\documentclass[usenatbib]{mnras}
\usepackage{graphicx}
\usepackage{verbatim}
\usepackage{color}
\usepackage{multirow}

\newcommand\kms{km$\,$s$^{-1}$}
\newcommand\Msol{M$_{\odot}$}

\begin{document}

\title[The ALFALFA HI mass function]{The ALFALFA HI mass function: A dichotomy in the low-mass slope and a locally suppressed `knee' mass}
\author[Jones et al.]{Michael G. Jones$^{1,2}$\thanks{E-mail: mjones@iaa.es}, Martha P. Haynes$^{1}$, Riccardo Giovanelli$^{1}$ \newauthor and Crystal Moorman$^{3}$
\\
$^{1}$Cornell Center for Astrophysics and Planetary Science, Space Sciences Building, Cornell University, Ithaca, NY 14853, USA
\\
$^{2}$Instituto de Astrof\'{i}sica de Andaluc\'{i}a, Glorieta de la Astronom\'{i}a s/n, 18008 Granada, Spain
\\
$^{3}$Department of Physics, Lynchburg College, 1501 Lakeside Drive, Lynchburg, VA 24501, USA}

\maketitle

\begin{abstract}
We present the most precise measurement of the $z = 0$ HI mass function (HIMF) to date based on the final catalogue of the ALFALFA (Arecibo Legacy Fast ALFA) blind HI survey of the nearby Universe. The Schechter function fit has a `knee' mass $\log (M_{*}\,h^{2}_{70}/\mathrm{M_{\odot}}) = 9.94 \pm 0.01 \pm 0.05$, a low-mass slope parameter $\alpha = -1.25 \pm 0.02 \pm 0.1$, and a normalisation $\phi_{*} = (4.5 \pm 0.2 \pm 0.8) \times 10^{-3} \; h^{3}_{70}\,\mathrm{Mpc^{-3}\,dex^{-1}}$, with both random and systematic uncertainties as quoted. Together these give an estimate of the HI content of the $z = 0$ Universe as $\Omega_{\mathrm{HI}} = (3.9 \pm 0.1 \pm 0.6) \times 10^{-4} \, h^{-1}_{70}$ (corrected for HI self-absorption). Our analysis of the uncertainties indicates that the `knee' mass is a cosmologically fair measurement of the $z = 0$ value, with its largest uncertainty originating from the absolute flux calibration, but that the low-mass slope is only representative of the local Universe. We also explore large scale trends in $\alpha$ and $M_{*}$ across the ALFALFA volume. Unlike with the 40 per cent sample, there is now sufficient coverage in both of the survey fields to make an independent determination of the HIMF in each. We find a large discrepancy in the low-mass slope ($\Delta \alpha = 0.14 \pm 0.03$) between the two regions, and argue that this is likely caused by the presence of a deep void in one field and the Virgo cluster in the other. Furthermore, we find that the value of the `knee' mass within the Local Volume appears to be suppressed by $0.18 \pm 0.04$ dex compared to the global ALFALFA value, which explains the lower value measured by the shallower HIPASS. We discuss possible explanations and interpretations of these results and how they can be expanded on with future surveys.
\end{abstract}

\begin{keywords}
radio lines: galaxies --- surveys
\end{keywords}

\section{Introduction}

The HI mass function, or HIMF, is the intrinsic distribution of galaxy HI masses in the Universe. Numerous HI surveys \citep{Rosenberg+2002,Zwaan+2005,Springob+2005,Martin+2010,Hoppmann+2015} have found this function to be well fit by a Schechter function \citep{Schechter1976}, a power law increasing towards lower masses and an exponential decline at high masses. The two shape parameters of this function are the power law exponent, usually referred to as the low-mass slope ($\alpha + 1$) on a logarithmic scale, and the `knee' mass ($M_{\ast}$) where the exponential decline begins.

Current cosmological simulations now include gas and baryonic processes \citep[e.g.][]{Vogelsberger+2014,Crain+2015} that alter the galaxy populations they produce. Many of the baryonic processes included in these models are functions of galaxy environment; for example they may depend on tidal stripping, the host halo mass, or the background UV radiation field. As HI typically has both a highly extended spatial distribution and is neutral, it is one of the most sensitive baryonic components to galaxy interactions and hard radiation fields. Therefore, the HIMF, and any variations with environment, are key constraints for the galaxy populations resulting from such simulations.

HIPASS \citep[HI Parkes All Sky Survey,][]{Barnes+2001,Meyer+2004}, which covered the entire Southern sky below declination $+2^{\circ}$ (21341 deg$^{2}$)\footnote{HIPASS' Northern extension \citep{Wong+2006} goes up to a declination of $+25.5^{\circ}$, however, this was not included in the HIPASS HIMF calculation and so is not considered in this paper.}, made one of the first robust measurements of the $z=0$ HIMF \citep{Zwaan+2005}, calculating a `knee' mass of $10^{9.86 \pm 0.03}$ \Msol \ and a low-mass slope of $-0.37 \pm 0.03$ (or $\alpha = -1.37 \pm 0.03$). The ALFALFA (Arecibo Legacy Fast ALFA\footnote{ALFA (Arecibo L-band Feed Array) is the name of the 7 beam feed horn array instrument with which the survey was performed.}) survey \citep{Giovanelli+2005} observed a smaller portion of the sky ($\sim$6900 deg$^{2}$) than HIPASS, but with substantially improved sensitivity, angular and velocity resolution, and depth. As Arecibo is located at a latitude of 18$^{\circ}$ N and can operate at a maximum zenith angle of about 20$^{\circ}$, the two surveys have minimal overlap. The ALFALFA 40 per cent ($\alpha$.40) HIMF was calculated by \citet{Martin+2010}, who found a marginally flatter slope ($\alpha = -1.33 \pm 0.02$) and higher `knee' mass ($10^{9.96 \pm 0.02}$ \Msol) than found in HIPASS.

Due to its much larger sky coverage the total survey volume of HIPASS is greater than that of ALFALFA, however ALFALFA's greater sensitivity actually results in a larger volume for the relevant portions of the HIMF. For example, a galaxy of HI mass $\log M_{\mathrm{HI}}/\mathrm{M_{\odot}} = 9$ and a velocity width of 100 \kms, would be detectable by ALFALFA out to a distance of approximately 80 Mpc \citep[based on the completeness limit defined in][]{Haynes+2011}, whereas the same galaxy would only be detectable out to about 35 Mpc in HIPASS \citep[at an equivalent completeness level,][]{Zwaan+2004}. This means that the maximum volume over which ALFALFA can probe the low-mass slope is about 4 times larger than what was probed by HIPASS. Similarly, at the HIMF `knee' mass, ALFALFA can probe out to almost 200 Mpc, whereas the equivalent distance for HIPASS is only 75 Mpc. Again, this means that the volume available in ALFALFA to study the `knee' mass is over 6 times greater than for HIPASS.\footnote{For more details on these calculations please see Appendix \ref{app:vol}.}

\citet{Zwaan+2005} also looked for environmental dependence of the HIMF, finding that the low-mass slope steepened slightly in high density environments, and that the `knee' mass was unaffected. However, as there was no optical galaxy survey covering the HIPASS sky that could be used to define environment, \citet{Zwaan+2005} defined environment based on the density of HIPASS sources themselves, therefore, these results are non-trivial to compare with \citep[see][for a more detailed discussion]{Jones+2016b}.

Using $\alpha$.40 \citet{Moorman+2014} assessed whether the shape of the HIMF was different in large cosmic voids compared to the walls that separate them. They found a small, but significant, decrease in the `knee' mass of void galaxies and a marginal flattening of the low-mass slope. \citet{Jones+2016b} used the neighbour density in optical and infrared surveys overlapping with the ALFALFA 70 per cent catalogue ($\alpha$.70) to look for trends based on a galaxy's local environment, finding an increase in the `knee' mass \citep[similar to][]{Moorman+2014} in higher density environments, but no significant change in the low-mass slope. 

With the final ALFALFA 100 per cent catalogue ($\alpha$.100) it is now possible to compare the HI properties of galaxies in two continuous regions spanning thousands of square degrees in the disparate environments covered by the Arecibo Spring and Fall skies. Thus, in this paper we take a different approach to the studies above and focus on the largest scale shifts in environment that we can, namely, the Spring and Fall sky regions of ALFALFA, inside and outside the Local Volume, and finally, proximity to the Virgo region. In addition, we present the most precise blind measurement of the global $z = 0$ HIMF to date along with a thorough analysis of its uncertainties. 

The paper is laid out as follows: Section 2 briefly describes the ALFALFA samples used throughout the paper, Section 3 outlines how the HIMF is estimated, and the resulting HIMF and its uncertainty estimates are presented in Section 4. Section 5 covers how the HIMF varies across the different sub-samples which we consider, and possible interpretations of these results are discussed in Section 6. Finally we draw our conclusions in Section 7. $H_{0}$ is assumed to be 70 $\mathrm{km\,s^{-1}\,Mpc^{-1}}$ throughout this paper.

\section{The ALFALFA sample}
\label{sec:alfalfa_sample}

The final ALFALFA footprint is approximately 6900 deg$^{2}$, or about a third the size of the HIPASS footprint. This area is split between two continuous regions in the directions of Virgo ($\sim$7.5 hr RA to $\sim$16.5 hr RA) and Pisces ($\sim$22 hr RA to $\sim$3 hr RA), with a declination range or approximately 0$^{\circ}$ to 36$^{\circ}$ in both cases. We will refer to these regions as the Arecibo ``Spring sky" and ``Fall sky" respectively, after the seasons in which they were observed. This area was observed blindly using a drift scan strategy that resulted in over $95$ per cent time efficiency, including all overheads. Both the data flagging and source extraction were carried out in part by automated processes \citep{Saintonge+2007}, but were ultimately completed by a person, who in the case of extragalactic sources would also identify the most likely optical counterpart from available overlapping surveys.

The typical rms noise level in ALFALFA is 2.4 mJy per beam (the ALFA beam is 3.3 by 3.8 arcmin at 21 cm) at 10 \kms \ resolution, compared to 13 mJy per beam (the Parkes beam diameter is 15.5 arcmin at 21 cm) at 26 \kms \ resolution for HIPASS. The respective redshift ranges of HIPASS and ALFAFLA are $-1280 \, \mathrm{km\,s^{-1}} < cz < 12700 \, \mathrm{km\,s^{-1}}$ and $-1600 \, \mathrm{km\,s^{-1}} < cz < 18000 \, \mathrm{km\,s^{-1}}$.

The ALFALFA 40 per cent catalogue \citep{Haynes+2011}, or $\alpha.40$, contained 40 per cent of the final footprint area, mostly in the Spring sky. The 100 per cent catalogue ($\alpha.100$) has now added significantly more area to the Fall sky as well as expanding the coverage in the Spring sky. These catalogues are available online at \url{http://egg.astro.cornell.edu/alfalfa/data/index.php} A summary of the $\alpha.100$ catalogue will be given in an upcoming paper (Haynes et al. 2018, in preparation). However, the catalogue description of $\alpha$.40 \citep{Haynes+2011} can be applied to the HI properties listed in $\alpha.100$, as we have done in this paper.

Estimating HI masses for ALFALFA sources requires first an estimate of their line-of-sight distance. To make this estimate ALFALFA uses the combination of a Local Volume flow model \citep{Masters2005}, group assignments based on the 2MRS \citep[2MASS Redshift Survey,][]{Huchra+2012,Crook+2007} and the catalogue of Nearby Optical Galaxies \citep[NOG,][]{Giuricin+2000,Springob2006}, as well as primary and secondary distances available in the literature. In addition, sources in the Virgo region are matched to the VCC \citep[Virgo Cluster Catalog;][]{Binggeli+1985} and assigned to the relevant clouds in the Virgo cluster. Further details of the distance estimation process are given in Section \ref{sec:results_a100}. With these distances the HI mass of the ALFALFA sources are then calculated using the standard equation:
\begin{equation}
\frac{M_{\mathrm{HI}}}{\mathrm{M}_{\odot}} = 2.356 \times 10^{5} D_{\mathrm{Mpc}}^{2} S_{21},
\end{equation}
where $D_{\mathrm{Mpc}}$ is the distance to the galaxy in Mpc and $S_{21}$ is its integrated flux in Jy \kms.

\begin{figure*}
\centering
\includegraphics[width=\textwidth]{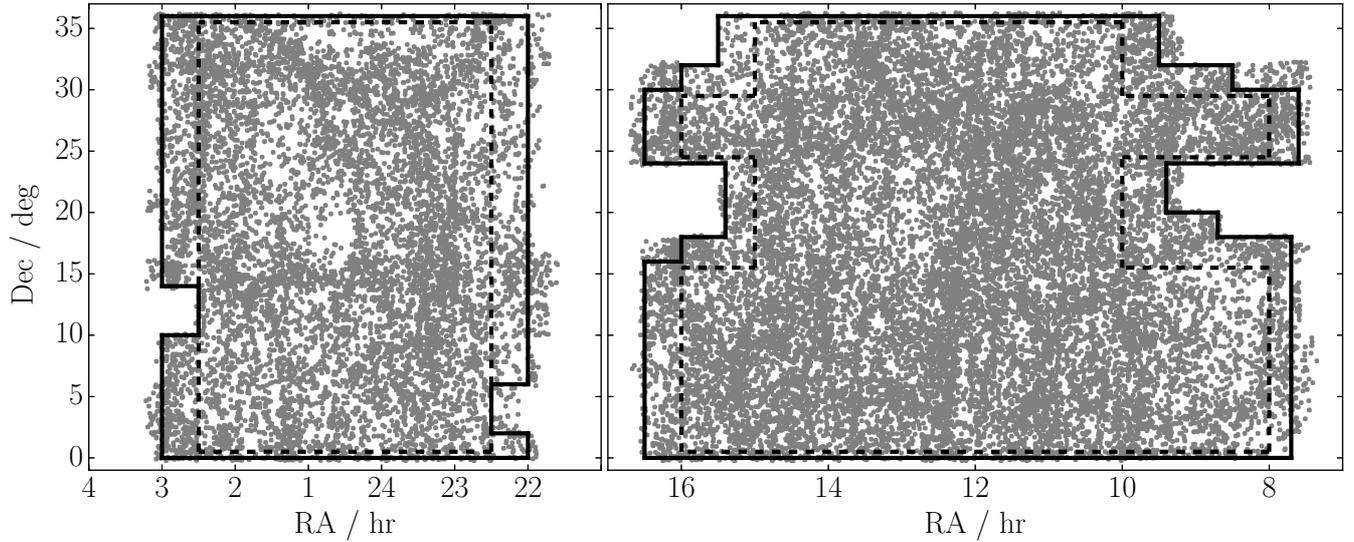}
\caption{The grey points show the sky positions of the high signal-to-noise extragalactic $\alpha$.100 sources with $v_{\mathrm{cmb}} < 15000$ \kms. The left side is the ALFALFA Fall sky region and the right side is the Spring sky region. The solid lines show the boundary which is used to calculate the HIMF and the dashed lines show the strict boundary that is used for comparison purposes. The strict boundary moves in much further in RA (in order to be very conservative) than in Dec because most of the variance in the boundary is determined by the time the observations started and stopped drift scanning, not by the top and bottom edges of the drift. The vertices of both boundaries are listed in Appendix \ref{sec:boundary}.}
\label{fig:skycov}
\end{figure*}

\section{Calculating HI Mass Functions}
\label{sec:HIMFcalc}

The HIMF represents the intrinsic number density of galaxies in the Universe as a function of their HI mass. This number density is usually denoted as
\begin{equation}
    \phi(M_\mathrm{HI}) = \frac{dN_{\mathrm{gal}}}{dV \: d\log_{10}(M_\mathrm{HI})},
\end{equation}
where $dN_{\mathrm{gal}}$ is the average number of galaxies in the volume $dV$, with HI masses that fall within a small logarithmic bin centred on $M_{\mathrm{HI}}$.

Converting the observed number counts (lower panel of Figure \ref{fig:a100_HIMF}) as a function of HI mass to the intrinsic number (Figure \ref{fig:a100_HIMF} upper panel) is a non-trivial process because ALFALFA is not a volume-limited sample and so the survey sensitivity must be corrected for. Furthermore, the sensitivity limit of an HI survey depends both on the source flux and velocity width. ALFALFA's sensitivity limits are explained in detail in \citet{Haynes+2011} and we use the 50 per cent completeness surface for Code 1 sources (signal-to-noise $>$ 6.5) as our sensitivity threshold throughout this paper.

Large scale structure (LSS) along the line-of-sight causes certain regions in redshift to be either sparsely or densely populated (relative to what would be expected in a uniform universe). Unless corrected for along with the survey sensitivity this can impact the form of the calculated HIMF. To account for these effects we use the 2-dimensional stepwise maximum likelihood (2DSWML) estimator \citep{Efstathiou+1988,Zwaan+2003,Zwaan+2005,Martin+2010}, also known as the $1/V_{\mathrm{eff}}$ method, which has been shown to be robust against the effects of LSS. For a detailed description of the application of this estimator to ALFALFA refer to \citet{Martin+2010} or \citet{Papastergis2013} Appendix A.

By assuming that the form of the HIMF is universal throughout the volume considered the distribution of sources along the line-of-sight drops out of the maximum likelihood process (and with it the HIMF normalisation), preventing adverse effects from LSS, but meaning that the normalisation constant must be estimated by other means. 

It has been demonstrated numerous times \citep[e.g.][]{Zwaan+2005,Martin+2010} that the HIMF is well fit by a Schechter function
\begin{equation}
\phi(M_\mathrm{HI}) = \ln(10) \: \phi_\ast \: \left( \frac{M_\mathrm{HI}}{M_\ast} \right)^{\alpha+1} \: e^{-\left( \frac{M_\mathrm{HI}}{M_\ast}\right)},
\end{equation}
where the fit parameters are $\phi_\ast$, the normalisation constant, $M_\ast$, the `knee' mass (we will often use $m_\ast = \log M_\ast / \mathrm{M_\odot}$), and the low-mass slope gradient ($\alpha + 1$). These parameters are used throughout this paper to describe the fits to the various HIMFs that we calculate.

An accurate definition is needed of the area on the sky which the survey is complete over. ALFALFA was conducted as a drift scan survey with the ALFA (Arecibo L-band Feed Array) instrument. Each observing night the telescope would be parked at a Dec between 0 and +36 degrees and take data as the sky drifted by: between RA 08:00 and 16:30 when observing the Spring part of the sky and between RA 22:00 and 03:00 when observing the Fall part. Each Dec strip was observed twice in this manner over the course of the survey, with the second observation offset by half a beam width. The vagaries of telescope time allocation are responsible for irregularities near the East and West edges of the sky box, which translated into occasionally poor or incomplete data coverage. To define the edges of the survey area we inspected the coverage maps by eye, and chose the RA and Dec extremes such that (as near as possible) the entire area has at least single pass coverage. 

The full $\alpha$.100 catalogue contains 25437 high signal-to-noise extragalactic sources. With the sky area trimmed to 6501 deg$^2$ (Figure \ref{fig:skycov}) the catalogue contains 24340 sources. The uncertainty in the determinations of this sky boundary and the impact that this may have on our results is assessed in Section \ref{sec:results_a100}. A final boundary is enforced in redshift space at a velocity of 15,000 \kms \ (relative to the Cosmic Microwave Background, CMB), beyond which ALFALFA suffers major incompleteness due to radio frequency interference. This further reduces the galaxy count to 23621. All the sub-samples discussed in the remainder of the paper were drawn from this sample of 23621 sources.\footnote{This does not include the samples with alternative distance estimates, for which the maximum velocity cut can change (Appendix \ref{sec:alt_dist}).}


\section{The ALFALFA HI mass function}
\label{sec:results_a100}

To calculate the global $\alpha$.100 HIMF we take the catalogue of sources described in the previous Section and impose the following cuts: $6 < \log M_{\mathrm{HI}}/\mathrm{M_\odot}$, $1.2 < \log W_{50}/\mathrm{km\,s^{-1}}$, and the 50 per cent completeness limit for high signal-to-noise sources \citep[][their equations 4 and 5]{Haynes+2011}. Note that we have intentionally avoided making a minimum distance cut to remove sources with high fractional distance uncertainties. The lowest mass bins do not appear particularly noisy so such a cut was deemed unnecessary, especially as it can also cause an artificial suppression of the lowest mass bins. As all ALFALFA sources are inspected by eye the separation between High Velocity Clouds (HVCs) and low recession velocity dwarf galaxies is expected to be highly reliable. While there does remain some possibility of sources being overlooked due to the bright foreground emission of HVCs, this is expected to be, at most, a handful of sources. Any such sources would also likely fall below the minimum HI mass cut. Furthermore, these ``mistaken identity" candidates are being followed up as part of the UCHVC (Ultra-Compact High Velocity Cloud) project \citep[e.g.][]{Adams+2013,Adams+2016}.

\begin{figure}
\centering
\includegraphics[width=\columnwidth]{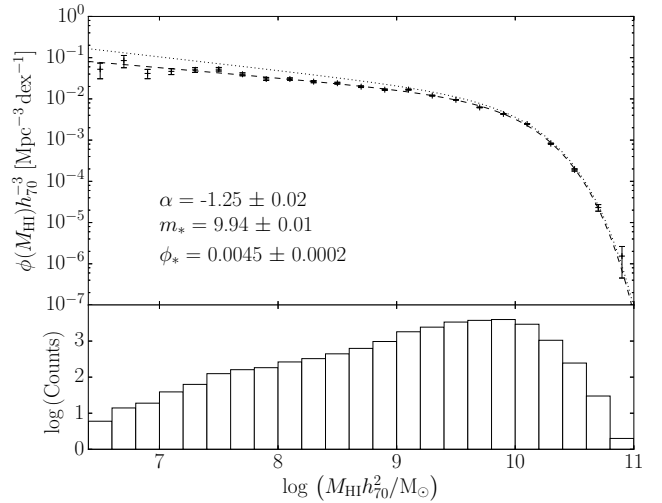}
\caption{The ALFALFA 100 per cent HIMF. The lower panel shows the number counts in logarithmic HI mass bins, and the points in the upper panel show the calculated intrinsic abundances after correcting for LSS and survey sensitivity using the 2DSWML method. The error bars and the corresponding fit errors displayed here are from Poisson counting errors only. A more detailed error analysis can be found in the text. The dashed line shows the best fit to the data and the corresponding Schechter function parameters are in the bottom left corner. The dotted line represents the ALFALFA 40 per cent best fit \citep{Martin+2010}.}
\label{fig:a100_HIMF}
\end{figure}

These further cuts give a final sample of 22831 galaxies from which the HIMF is computed using the $1/V_{\mathrm{eff}}$ method. A Schechter function is then fit in linear space to the non-parametric result (shown in Figure \ref{fig:a100_HIMF}) giving the parameters as $\alpha = -1.25 \pm 0.02$, $m_{*} = 9.94 \pm 0.01$, and $\phi_{*} = 4.5 \pm 0.2 \times 10^{-3} \; \mathrm{Mpc^{-3} \, dex^{-1}}$, where the fit errors quoted here are due only to the Poisson errors and do not include contributions from distance uncertainty or other random and systematic effects. The following subsections will discuss a number of other sources of error and how we estimate them. Our estimates are collated in table \ref{tab:errors}.

\subsection{Random flux and distance errors}
\label{sec:dist_errs}

In order to estimate the error introduced due the uncertainty in source distances we took a Monte Carlo approach, creating many realisations of the HIMF calculation, altering the sources distances (and therefore masses) each time. There are seven methods that are used to calculate the distance to ALFALFA sources (the assignment occurs in the order listed, top given highest preference):
\begin{itemize}
    \item Literature primary distances are used for all sources that are associated with an optical counterpart for which such a measurement exists.
    \item Sources that are associated with counterparts in the Virgo Cluster Catalog \citep{Binggeli+1985} are assigned the distance to the relevant Virgo cloud \citep[as described in][]{Hallenbeck+2012}.
    \item Literature secondary distances are used for sources with $v_{\mathrm{cmb}} < 6000$ \kms \ that are associated with optical counterparts for which such a measurement exists. 
    \item Galaxies assigned to a group are given the mean recession velocity of the group. For $v_{\mathrm{cmb}} < 6000$ \kms \ a flow model \citep{Masters2005} is used to estimate the distance.
    \item For group galaxies with $v_{\mathrm{cmb}} > 6000$ \kms \ pure Hubble flow (with $H_{0} = 70 \; \mathrm{km \, s^{-1} \, Mpc^{-1}}$) is assumed.
    \item The remaining sources with $v_{\mathrm{cmb}} < 6000$ \kms \ are assigned distances from the flow model.
    \item The remaining sources with $v_{\mathrm{cmb}} > 6000$ \kms \ are assigned distances using Hubble flow.
\end{itemize}
To account for the uncertainty in the distances to ALFALFA sources we followed a similar procedure to that of \cite{Papastergis+2016}. We ran 1000 iterations of the HIMF calculation, each time randomly adjusting the distance to each source as follows:
\begin{itemize}
    \item Primary distances and assignments to Virgo are assumed to have a 10 per cent Gaussian uncertainty.
    \item Secondary distances are drawn from a Gaussian distribution centred on the initial value and with a width of 20 per cent of that value.
    \item Galaxies assigned to groups have a Gaussian uncertainty introduced due to the uncertainty in $H_{0}$, which we take to be 3 $\mathrm{km\,s^{-1}\,Mpc^{-1}}$, as well as the inclusion of a group peculiar velocity that is assigned to all members of the group (drawn from a Gaussian distribution centred on 0 and with a width of 300 \kms).
    \item The remaining sources are given a Gaussian error based on the quadrature sum of the uncertainty in $H_{0}$ and the typical value of galaxy peculiar velocities, which we take to be 160 \kms.
\end{itemize}

In addition to the distance uncertainties, the impact of flux measurement uncertainty was introduced to each source by adding a random Gaussian error with a width of the quoted flux uncertainty in the $\alpha$.100 catalogue. The output Schechter parameter values of the 1000 resulting iterations were then fit with normal distributions, giving standard deviations of $\sigma_{\alpha} = 0.007$, $\sigma_{m_\ast} = 0.004$, and $\sigma_{\phi_\ast} = 7 \times 10^{-5}  \; \mathrm{Mpc^{-3} \, dex^{-1}}$. Combining these in quadrature with the uncertainty in an individual fit (from the Poisson counting errors) gives an estimate of the total random error in the parameter values: $\alpha$.100 as $\alpha = -1.25 \pm 0.02$, $m_{*} = 9.94 \pm 0.01$, and $\phi_{*} = 4.5 \pm 0.2 \times 10^{-3} \; \mathrm{Mpc^{-3} \, dex^{-1}}$.

\subsection{Choice of boundary}

The ALFALFA survey does not have perfectly binary coverage, and therefore defining the area in which the HIMF will be estimated is somewhat subjective. The survey was performed using a double pass drift-scan strategy and thus the edges are ragged in RA due to differing start time. The nominal boundary was chosen to ensure the full area had at least single coverage (as far as was possible). A very conservative boundary was also created by moving 0.5$^{\circ}$ in Dec and to the nearest half hour of RA inside the nominal boundary (see Figure \ref{fig:skycov}). Enforcing this highly conservative boundary cut changes the best fit parameters to $\alpha = -1.26 \pm 0.02$, $m_{*} = 9.94 \pm 0.01$, and $\phi_{*} = 4.8 \pm 0.2 \times 10^{-3} \; \mathrm{Mpc^{-3} \, dex^{-1}}$, indicating that for $\alpha$ the uncertainty in the choice of boundary is a similar scale source of error as the Poisson counting uncertainty. 

It should be noted, however, that the patchy coverage at the extremes of the survey footprint are unlikely to be the main cause of this shift in the parameter values. Altering the boundary in this way creates a change in two inseparable systematic effects. The first is the intended effect, the elimination of the non-uniform coverage near the edge of the survey. The second is that the footprint of the survey is changed, which means that it covers a slightly different range of local environments. The strict boundary chosen removes area from the edges of the Fall and Spring skies, but not from the vicinity of Virgo, which (as will be discussed below) will act only to steepen the low-mass slope, meaning that the above approach likely gives an overestimate of the impact of the choice of boundary. If no boundary is imposed $\alpha$ and $M_{*}$ are unchanged from the values with the nominal boundary ($\phi$ cannot be calculated without knowing the sky area).

\subsection{Cosmic and sample variance}

Cosmic variance is a phrase used to mean a number of different things in different contexts, here we use it to mean that the Universe contains regions of different underlying density, which may potentially impact the form of the mass function (MF). Therefore, unless a survey volume is large enough to contain a fair sample of all of these environments there will be a systematic bias in the resulting MF, regardless of how precisely it can be determined within that volume, and therefore it will not be representative of the Universe's global MF (at that redshift). To estimate the scale of this effect we simply take the difference in the HIMFs in the ALFALFA Spring and Fall regions. These are two regions with completely disparate large scale environments and so should give a reasonable approximation of how much the parameters can be expected to vary on larger scales (see Section \ref{sec:spring_v_fall}).

There is another effect of cosmic variance (which here we will label ``sample variance"), which is that it limits how well the HIMF can be determined inside the survey volume because the HIMF may vary across it. To estimate this effect we split the survey area into 46 approximately equal area, contiguous regions and jackknifed the HIMF calculation, removing one region each time. The sums of the squared deviations from the mean parameter values (across all jackknife samples), multiplied by $(N-1)/N$, give estimates of the parameter uncertainties due to sample variance. These are: $\sigma_{\alpha} = 0.02$, $\sigma_{m_\ast} = 0.01$, and $\sigma_{\phi_\ast} = 3 \times 10^{-4}  \; \mathrm{Mpc^{-3} \, dex^{-1}}$.

\subsection{Flow model uncertainty}

ALFALFA uses a local Universe flow model \citep{Masters2005} and additional corrections (see Section \ref{sec:dist_errs} and Appendix \ref{sec:alt_dist}) to estimate source distances. In Section \ref{sec:dist_errs} we test the impact of random errors within this framework, but that does not give an estimate of the systematic uncertainty stemming from the use of flow model itself, as opposed to some other method for estimating distances. As pointed out in \citet{Masters+2004}, an incorrect flow model can lead to substantial systematic biases in the derived HIMF. To estimate the scale of this uncertainty we recalculate the HIMF using both the flow model of \citet{Mould+2000} and pure Hubble flow only (discussed further in Appendix \ref{sec:alt_dist}). The resulting Schechter function parameters are shown in Table \ref{tab:param_vals}. By taking the standard deviation between these three measurements we obtain a very approximate estimate of the uncertainty due to the distance model we use. These are: $\sigma_{\alpha} \approx 0.01$, $\sigma_{m_\ast} \approx 0.01$, and $\sigma_{\phi_\ast} \approx 1 \times 10^{-4}  \; \mathrm{Mpc^{-3} \, dex^{-1}}$.

\subsection{Absolute flux scale}

Perhaps the largest source of systematic error for $M_{*}$ is the calibration of the absolute flux scale. The flux scale is calibrated by observing a calibration diode intermittently during drift-scans and an additional correction is applied during the data reduction process after matching the continuum source fluxes in the survey area to known catalogues \citep[][and references therein]{Haynes+2011}. This calibration is estimated to be accurate to about 10 per cent, which corresponds to a systematic uncertainty of 0.04 dex in $M_{*}$. However, this uncertainty should not impact the low-mass slope as it acts to shift all sources equally, simply creating a shift in the horizontal direction in the HIMF. 

As discussed in detail in \citet{vanZee+1997}, with carefully constructed observations a flux calibration at the level of a few per cent can be obtained, however, this rises to about 5 per cent when baseline uncertainties are included. The estimate for ALFALFA is larger than this because in a blind survey each observation cannot be tailored with the goal of producing high procession fluxes \citep[as was the case for][]{vanZee+1997}, and because it is intended to be conservative, including various sources of uncertainty such as drift in the system gain, standing waves due to interference or continuum emission, and baseline variations. Further details of ALFALFA's calibration can be found in Section 5.2 of \citet{Haynes+2011}.

It should also be noted that ALFALFA's source extraction procedure is non-optimal for sources that are larger than the Arecibo primary beam, therefore, the largest galaxies (in angular extent) may have larger flux uncertainties. However, making the simple approximation that the HI radius of a galaxy is double its optical radius (optical radii taken from the Arecibo General Catalog, a galaxy database maintained by MPH and RG) we estimate that only 2 per cent of ALFALFA detections have HI extents greater than 4', thus, this is not a major concern.

\subsection{The 2DSWML method}

The final source of uncertainty we consider is bias in the 2DSWML method itself. To do this we created 100 mock catalogues of approximately 1 million HI sources each \citep[following the methodology of][to create the mocks]{Jones+2015}, all with the same input HIMF. The large source count was required such that Poisson errors were not a limiting factor. The HIMF for each mock was calculated using the same process as for the real ALFALFA data. As before, the resulting distributions of the Schechter function fit parameters were approximated by Gaussians and gave the uncertainties as: $\sigma_{\alpha} = 0.02$, $\sigma_{m_\ast} = 0.002$. $\sigma_{\phi_\ast}$ was not calculated in this analysis as the 2DSWML method does not return the normalisation.

\subsection{HI self-absorption}

In principle 21 cm emission can be absorbed by other HI residing in the source galaxy, this is known as HI self-absorption. This effect is often neglected entirely because the cross-section for interaction between an HI atom and a 21 cm photon is very small, and it is a challenging and somewhat controversial topic, with some authors finding that all but the most highly inclined galactic discs are almost transparent to HI \citep[e.g][]{Giovanelli+1994}, while others indicate that typical corrections could be as high as 30 per cent for all galaxies \citep[e.g.][]{Braun+2009,Braun2012}. To assess what impact this has on ALFALFA, and in particular our estimate of $\Omega_{\mathrm{HI}}$, we follow a procedure based on \citet{Giovanelli+1994} using the HI data from ALFALFA in combination with the exponential radii and axial ratios calculated in SDSS DR14 \citep{SDSSDR14}. 

If a single galaxy could be rotated and viewed from multiple orientations then as it became more edge-on its total integrated HI emission would be expected to decrease somewhat as the line-of-sight takes a longer path through the disc. By using axial ratios as a proxy for the inclinations of many galaxies (instead of rotating a single galaxy) the rate at which the \textit{average} observed HI mass changes with inclination can be determined. For a sample of galaxies with similar morphologies and HI masses, the average observed HI mass should be independent of inclination, except for self-absorption.

As the vast majority of ALFALFA sources are late-type spiral galaxies we will assume the morphology has a minimal effect. However, the intrinsic axial ratio of galaxies is likely dependent on their mass, so we will exclude dwarf galaxies from this analysis. In addition, it is necessary to remove sources where the minor axis is comparable in angular size to the resolution of the optical image. However, making almost any cut to the sample based on the optical properties is problematic. For example, if this cut is made just at a constant angular resolution then it is creating a distance-dependent effect, but if it is scaled linearly with distance then it creates a very stringent resolution requirement at small distances. We therefore decided (starting from the 22831 high signal-to-noise sources used to calculate the $\alpha$.100 HIMF) only to consider sources in the distance range $100 < D/{\mathrm{Mpc}} < 150$, which effectively sets the minimum HI mass as $\log M_{\mathrm{HI}} \sim 9.1$, we also only consider sources that have unambiguously identified counterparts that are ``primary'' objects in SDSS, and finally we require that the major axis exponential radius (in $r$-band) is greater than $10 \times \frac{100 \, \mathrm{Mpc}}{D}$ arcsec---with this value even the most inclined galaxies have minor radii of $\sim$1 arcsec, thus the measured axial ratio should not be strongly influenced by the image resolution. This leaves a sample of 2022 galaxies. While this selection undoubtedly results in some level of bias with respect to inclination (axial ratio), it is aimed at minimising those sources of bias of which we are aware.

With this sample we fit a linear trend, using weighted least squares regression, between $\log (a/b)$ and $\log M_{\mathrm{HI}}$. \citet{Giovanelli+1994} split their sample into distance bins in order to minimise distance-dependent biases, however, ALFALFA is a considerably better characterised population (from the HI perspective) than their sample was, hence, we instead weight each source by the factor $M_{\mathrm{HI}}/V_{\mathrm{eff}}$ (with the $V_{\mathrm{eff}}$ values truncated to match the distance boundaries above). This is essentially weighting by mass density, which should produce a correction factor that is appropriate for correcting $\Omega_{\mathrm{HI}}$. 

We note here that without any weighting the sample is consistent (within 1-$\sigma$) with galactic discs being entirely transparent to HI emission, except in cases where the disc is almost exactly edge-on. With the mass density weighting the correction factor is found to be $\Delta \log M_{\mathrm{HI}} = (0.13 \pm 0.03) \log (a/b)$, which is consistent with that found by \citet{Giovanelli+1994}. An increase in the dependence between $\log (a/b)$ and $\log M_{\mathrm{HI}}$ is expected when the data are weighted by their $V_{\mathrm{eff}}$ values as sources with high velocity widths (i.e. those that are more inclined) will be underrepresented in the unweighted data due to the selection effects of the survey. Using this relation the most highly inclined galaxies in the sample require a correction of about 35 per cent, whereas at the log--mean axial ratio ($\log (a/b) = 0.36$) the correction is $11 \pm 3$ per cent, which we will use as our estimate of the correction required for $\Omega_{\mathrm{HI}}$. 

We do not use this value to make any correction to the shape or normalisation of the HIMF itself. While this correction was designed to be appropriate for the measurement of the total cosmic HI mass density, the correction likely changes across the mass range probed by ALFALFA. Given the difficulties in creating a relatively unbiased sample for the massive galaxies in the survey, we make no attempt to quantitatively estimate the correction to the HIMF's shape due to self-absorption, but we note that it is reasonable to expect the `knee' mass to also increase by $\sim$10 per cent as this region of the HIMF makes the dominant contribution to $\Omega_{\mathrm{HI}}$, and that corrections to the low-mass slope are likely quite small, as we expect the self-absorption correction to both decrease with decreasing mass and to do so gradually.

This result indicates that galactic discs are mostly transparent to HI emission, but a small correction is necessary to avoid bias. However, this analysis only considers effects that scale with inclination. \citet{Braun2012} suggests that a large fraction of the HI in galactic discs may reside in high density clouds (of $\sim$100 pc in diameter) which could result in approximately 30 per cent of HI emission being self-absorbed. If these finding are correct then most of this self-absorption would not be apparent in an anaylsis such as ours because inclination would not be relevant for self-absorption occurring within such compact clouds. Having said this, those results are based on only 3 galaxies and rely on the accuracy of the modelling of these dense clouds. To really resolve this issue will require extremely deep, $\sim$100 pc scale HI imaging of a large sample of galaxies, which is not feasible with existing facilities.

\begin{table*}
\centering
\caption{The different contributions to both random and systematic errors in the calculated values of the Schechter function fit to the HIMF. Poisson errors correspond to the Schechter function fit uncertainty due to the finite number of sources contained in each bin of the non-parametric HIMF calculated by the 2DSWML method. The random flux and distance errors are from the uncertainty in measuring a source's flux from its spectrum and from determining its distance from its position and heliocentric velocity. The boundary error is a conservative estimate of the effect of small areas without complete coverage near the edges of the survey. This source of error is impossible to separate from environmental dependence of the HIMF and most of this contribution is likely due to the removal or inclusion of more area covering a given environment, rather than from the effects of non-uniform coverage itself. The cosmic variance term corresponds to our estimate of how much the HIMF may differ in a volume larger than ALFALFA's. Again this is a very conservative estimate taken as the difference between the disparate ALFALFA Spring and Fall skies. Sample variance (as we have called it) is really another form of cosmic variance, but on a scale smaller than the survey volume. This was estimated by jackknifing the sample across 46 approximately equal area, contiguous regions. The systematic distance error due to the choice of flow model was estimated by considering three different possible models. There is also another source of error from the flux estimates which corresponds to the absolute scale of the observations, this is estimated to be accurate to about 10\%. Finally, the 2DSWML or $1/V_{\mathrm{eff}}$ method has an intrinsic error associated with it, which we estimated through simulated datasets.}
\label{tab:errors}
\begin{tabular}{lcccccccc}
\hline
\hline
\multirow{2}{*}{Parameter} & \multicolumn{2}{c}{Random uncertainties} & \multicolumn{6}{c}{Systematic uncertainties}                                                         \\
                           & Poisson             & Flux \& Dist       & Boundary           & Cosmic Var.             & Sample Var.        & Dist. Model & Abs. Flux & 2DSWML \\ \hline
$\alpha$                                         & 0.01                & 0.007              & $\sim$0.01               & $\sim$0.1               & 0.02               & $\sim$0.01        & -         & 0.02   \\
$m_{\ast}  + 2\log h_{70}$                       & 0.01                & 0.004              & $\sim$0.01               & $\sim$0.02              & 0.01               & $\sim$0.01        & $\sim$0.04      & 0.002  \\
$\phi_{*} / h^{3}_{70} \, \mathrm{Mpc^{-3} \, dex^{-1}}$&$2 \times 10^{-4}$  & $7 \times 10^{-5}$ & $\sim 3 \times 10^{-4}$ & $\sim 6 \times 10^{-4}$ & $3 \times 10^{-4}$ & $\sim 1 \times 10^{-4}$        & -         & -      \\ \hline
\end{tabular}
\end{table*}

\subsection{Final estimates of the HIMF and  $\Omega_{\mathrm{HI}}$}

Adding the random and systematic effects from table \ref{tab:errors} separately in quadrature (with the exception of cosmic variance) gives the final estimate of the HIMF within the ALFALFA volume as: $\alpha = -1.25 \pm 0.02 \pm 0.03$, $m_{*} = 9.94 \pm 0.01 \pm 0.04$, and $\phi_{*} = (4.5 \pm 0.2 \pm 0.4) \times 10^{-3} \; \mathrm{Mpc^{-3} \, dex^{-1}}$, where the first quoted errors correspond to our combined estimates of the 1-$\sigma$ random uncertainties and the second to the systematic uncertainties. This indicates that for sample sizes of $\sim$25000 or fewer, the Poisson uncertainties are reasonable estimates of the random uncertainties, but the systematic uncertainties are likely significantly larger. The inclusion of the uncertainty of cosmic variance gives an estimate of how well ALFALFA can constrain the global $z = 0$ form of the HIMF: $\alpha = -1.25 \pm 0.02 \pm 0.1$, $m_{*} = 9.94 \pm 0.01 \pm 0.05$, and $\phi_{*} = (4.7 \pm 0.2 \pm 0.8) \times 10^{-3} \; \mathrm{Mpc^{-3} \, dex^{-1}}$.

The value of $\Omega_{\mathrm{HI}}$ (the HI content of the Universe) can be estimated from the HIMF by integrating the Schechter function, which gives
\begin{equation}
    \Omega_{\mathrm{HI}} = \frac{1}{\rho_{\mathrm{c}}} \phi_\ast M_\ast \Gamma(\alpha+2),
\end{equation}
where $\rho_{\mathrm{c}}$ is the Universe's critical density (calculated assuming $H_{0} = 70 \; \mathrm{km \, s^{-1} \, Mpc^{-1}}$). The limits of the integral are taken to be indefinite because both the extreme low and high mass regimes make negligible contributions. This gives $\Omega_{\mathrm{HI}} = 3.9 \pm 0.1 \pm 0.6 \times 10^{-4}$ which is consistent at the 2-$\sigma$ level (random) with the value from $\alpha$.40 \citep{Martin+2010}, when the scaling for HI self-absorption is removed.
\footnote{The systematic uncertainty in $\Omega_{\mathrm{HI}}$ is difficult to measure because it requires estimates of the covariance between the systematic uncertainties in $\alpha$, $m_{\ast}$, and $\phi_{\ast}$, which are not available to us. Treating all the systematic uncertainties estimated above as independent would likely lead to an over estimate of systematic uncertainty in $\Omega_{\mathrm{HI}}$ ($1.0 \times 10^{-4}$), while treating them as highly correlated would likely lead to an underestimate ($0.1 \times 10^{-4}$). We therefore choose only to consider the most dominant source of uncertainty, the impact of the absolute flux calibration of $m_{\ast}$, which produces the quoted level of uncertainty ($0.6 \times 10^{-4}$). As this was a conservative estimate to begin with and it is the dominant source of uncertainty for $\Omega_{\mathrm{HI}}$, it is a reasonable estimate of the total systematic uncertainty.}
Our value is $\sim$15 per cent lower because both the `knee' mass is lower and the low-mass slope is flatter in $\alpha$.100 than $\alpha$.40, both of which act to decrease $\Omega_{\mathrm{HI}}$. This value is completely consistent (again, with the self-absorption correction removed) with that of HIPASS \citep{Zwaan+2005}, however, this agreement hides the substantially different shapes of the HIMF, which do not agree within the random errors. The $\Omega_{\mathrm{HI}}$ values are consistent because our `knee' mass is larger but our low-mass slope is substantially flatter.

\citet{Driver+2010} estimated that volumes larger than $10^{7}$ Mpc$^{3}$ with square survey footprints will have minimal contributions from cosmic variance. The volume over which we calculate the $\alpha$.100 HIMF is approximately half of this value, and that volume is split into two separate pieces (Figure \ref{fig:skycov}) which acts to reduce the volume needed to overcome cosmic variance. This suggests that for sources which are detectable over almost all this volume ($M_\ast$ galaxies) our Schechter function parameter values should only be weakly impacted by cosmic variance, however, for low-mass galaxies the accessible volume is only $\sim$10$^{5}$ Mpc$^{3}$, which may be strongly biased by cosmic variance. This is consistent with our estimates of cosmic variance in table \ref{tab:errors}. In other words, while the `knee' mass of the ALFALFA HIMF should be viewed as cosmologically fair, the low-mass slope should be considered a measurement of the value in the nearby Universe, and it therefore may not be representative of the volume beyond the Local Supercluster.

\begin{table*}
\centering
\caption{Schechter function parameter values for each of the ALFALFA samples for which the HIMF was calculated. Note that maximum recession velocities (relative to the CMB) are enforced as derived distance cuts at $v_{\mathrm{max}}/H_0$. The error estimates listed account only for Poisson counting uncertainties, see the text for estimates of the contributions of other error sources. The Spring and Fall samples represent the HIMF in the two separate regions of the ALFALFA survey in the Arecibo Spring and Fall skies. The strict bound samples have highly conservative boundaries that eliminate portions of the survey area that might not have complete coverage, this also changes the area considered which creates a shift in the parameters due to local cosmic variance. The two Virgo samples are for within 3 Mpc of the cluster centre (M87) and a wide slice covering the full ALFALFA Dec range and 1 hr in RA. Finally, the Hubble and \citet{Mould+2000} calculations are based on two alternative methods for calculating the source distances.}
\label{my-label}
\begin{tabular}{| l | c | c | c | c | c |}
\hline \hline
Sample              & $v_{\mathrm{max}} / \mathrm{km\,s^{-1}}$ & $N_{\mathrm{gal}}$ & $\alpha$ & $m_{*} + 2\log h_{70}$ & $\phi_{*} / h^{3}_{70} \, \mathrm{Mpc^{-3} \, dex^{-1}}$          \\  
\hline
$\alpha.100$                  & 15000                                            & 22831                      & $-1.25 \pm 0.02$  & $9.94 \pm 0.01$  & $4.5 \pm 0.2 \times 10^{-3}$ \\
$\alpha.100$ Spring           & 15000                                            & 14391                      & $-1.29 \pm 0.02$  & $9.94 \pm 0.02$  & $4.9 \pm 0.3 \times 10^{-3}$ \\
$\alpha.100$ Fall             & 15000                                            & 8440                       & $-1.15 \pm 0.02$  & $9.92 \pm 0.02$  & $4.3 \pm 0.3 \times 10^{-3}$ \\
$\alpha.100$ Strict bound     & 15000                                            & 19268                      & $-1.26 \pm 0.02$  & $9.94 \pm 0.01$  & $4.8 \pm 0.2 \times 10^{-3}$ \\
$\alpha.100$ Spring Strict bound     & 15000                                     & 12318                      & $-1.30 \pm 0.02$  & $9.94 \pm 0.02$  & $5.0 \pm 0.3 \times 10^{-3}$ \\
$\alpha.100$ Fall Strict bound     & 15000                                       & 6950                       & $-1.15 \pm 0.03$  & $9.91 \pm 0.02$  & $4.8 \pm 0.3 \times 10^{-3}$ \\
$\alpha.100a$                 & 4000                                             & 3815                       & $-1.22 \pm 0.02$  & $9.76 \pm 0.04$  & $6.2 \pm 0.5 \times 10^{-3}$ \\
$\alpha.100a$ Spring          & 4000                                             & 2634                       & $-1.24 \pm 0.02$  & $9.75 \pm 0.04$  & $6.2 \pm 0.6 \times 10^{-3}$ \\
$\alpha.100a$ Fall            & 4000                                             & 1181                       & $-1.08 \pm 0.03$  & $9.70 \pm 0.05$  & $7.7 \pm 0.8 \times 10^{-3}$ \\
Virgo RA slice               & 3000                                              & 695                        & $-1.23 \pm 0.05$  & $9.60 \pm 0.10$  & $2.6 \pm 0.6 \times 10^{-2}$ \\
Extended Virgo cluster        & -                                                & 272                        & $-1.20 \pm 0.12$  & $9.54 \pm 0.26$  & $0.30 \pm 0.16$ \\
$\alpha.100$ (Hubble)         & 15000                                            & 22815                      & $-1.25 \pm 0.01$  & $9.94 \pm 0.01$  & $4.6 \pm 0.1 \times 10^{-3}$ \\
$\alpha.100$ (Mould et al.)   & 15000                                            & 22693                      & $-1.26 \pm 0.02$  & $9.96 \pm 0.01$  & $4.3 \pm 0.2\times 10^{-3}$ \\
\hline
\end{tabular}
\label{tab:param_vals}
\end{table*}

\section{Variations in the HIMF}

We have found that the HIMF varies across the ALFALFA volume. In this Section we present these large scale changes in the shape of the HIMF across the two regions of the survey, the nearby and full samples, and in the direction of the Virgo cluster.

\subsection{Comparison of the Spring and Fall sky}
\label{sec:spring_v_fall}

\begin{figure}
\centering
\includegraphics[width=\columnwidth]{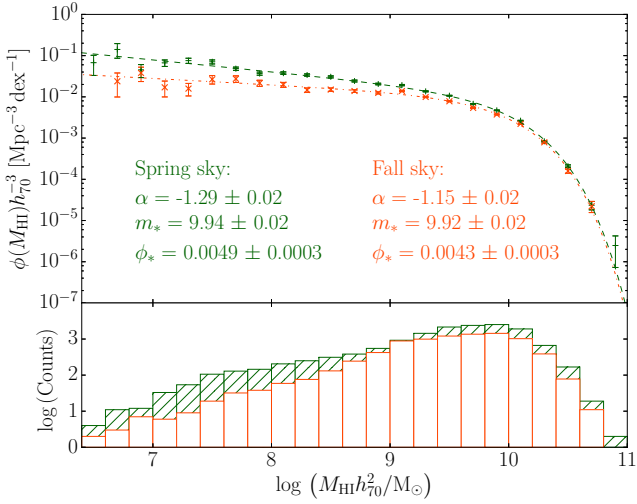}
\caption{The HIMFs of the $\alpha$.100 Spring and Fall skies. The lower and upper panels are as previously described, with the green plus signs, dashed line and filled bars corresponding to the Spring sky, and the orange crosses, dotted line and unfilled bars to the Fall sky. The `knee' mass in the two directions agrees within the uncertainties, but the Fall sky has a significantly flatter low-mass slope.}
\label{fig:a100_HIMF_SF}
\end{figure}

The nearby LSS in the Spring and Fall portions of the ALFALFA footprint is decidedly disparate, with the Virgo cluster and the Local Supercluster dominating the Spring direction, while the Fall sky is sparsely populated due to the deep void in the foreground of the Pisces-Perseus supercluster (PPS). Contrasting the HIMF in these two directions is the first step to assessing how it may vary between large scale over and under densities. To make the comparison the $\alpha$.100 volume is simply split into the Spring and Fall directions, and the HIMF calculated in each independently. Figure \ref{fig:a100_HIMF_SF} shows the two HIMF and their Schechter function fits.

Separating the population in this way immediately makes it clear that the Fall region has a substantially flatter low-mass slope than the Spring region, at over 3-$\sigma$ significance (based on Poisson random uncertainties), whereas the `knee' masses in the two directions agree within their errors. This difference in the slope persists in the nearby Spring and Fall catalogues at similar significance, as the Fall low-mass slope is somewhat lower in the nearby volume (see table \ref{tab:param_vals}).

\subsection{Comparison of interior and exterior to 4000 \kms}

\begin{figure}
\centering
\includegraphics[width=\columnwidth]{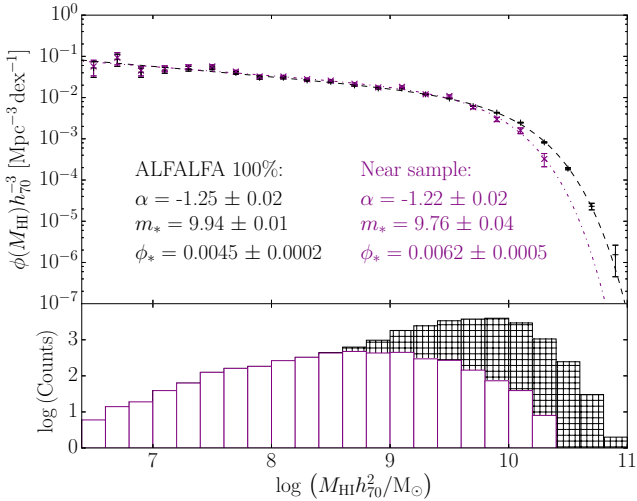}
\caption{The HIMFs of the $\alpha$.100a ($v_\mathrm{cmb} < 4000$ \kms) and full $\alpha$.100 samples. The lower panel shows the observed counts for the near (unfilled) and full samples (cross hatching), while the upper panel shows the inferred HIMF of those samples with purple crosses and black plus signs, respectively. The Schechter function fits are shown with dash-dot (near) and dashed (full) lines. These two samples appear to have best fit `knee' masses that are different by almost 0.2 dex, but similar low-mass slopes.}
\label{fig:a100_HIMF_near}
\end{figure}

To test of large scale effects with the line-of-sight distance within the ALFALFA volume we compare the $\alpha$.100 HIMF within $\sim$57 Mpc (4000 \kms$/H_{0}$) to that of the full sample. This nearby sample will be referred to as $\alpha$.100a. At this distance the ALFALFA 50 per cent completeness limit would just include a source of $\log M_{\mathrm{HI}}/\mathrm{M_\odot} = 8.5$ and a velocity width of 50 \kms. Therefore, this represents a reasonable cutoff beyond which few objects on the low-mass slope can be detected. Figure \ref{fig:a100_HIMF_near} shows that the low-mass slopes of the nearby sample and the whole $\alpha$.100 are within 1-$\sigma$ (Poisson) agreement, but that the `knee' mass is locally lower by 0.18 dex. The former is unsurprising because (by construction) the nearby sample contains virtually all the objects detected on the low-mass slope and therefore much below $\log M_{\mathrm{HI}} = 10^{9}$ \Msol \ the data points become identical. The latter is less straightforward to explain, with the difference in $m_\ast$ being at approximately the 4-$\sigma$ (Poisson) level.

\subsection{The HI mass function of the Virgo cluster}

\begin{figure}
\centering
\includegraphics[width=\columnwidth]{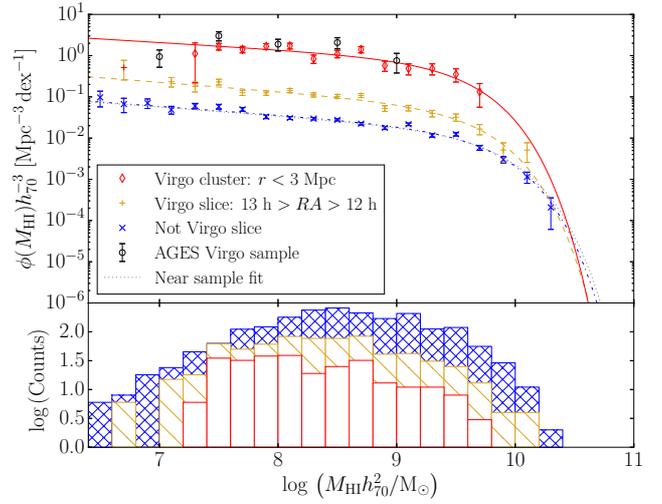}
\caption{The HIMF of the Virgo region compared to that of the $\alpha$.100a sample. Again the lower panel shows the raw numbers counts in each mass bin. The red, unfilled bars are the extended Virgo cluster sample, the single hatched gold bars are the Virgo slice sample, the blue cross hatched bars are the $\alpha$.100a sample minus the Virgo slice, and the black circles are our estimates of the AGES Virgo HIMF (with simplifying assumptions, see Section \ref{sec:discuss_virgo}). The upper panels shows the non-parametric HIMF bin values and the Schechter function fits to each.}
\label{fig:a100_HIMF_virgo}
\end{figure}

Defining the volume that encompasses the Virgo cluster is non-trivial and there is no single definition, we therefore take two approaches. First, we take simply a box in RA (between 13 and 12 hr) and Dec (between 0$^{\circ}$ and 36$^{\circ}$) that stretches out to a maximum distance of $\sim$43 Mpc (3000 \kms$/H_{0}$), as a very crude approximation of the greater Virgo volume, including all the major clouds and filaments that surround the cluster itself. This region contains 695 sources and its HIMF is shown in Figure \ref{fig:a100_HIMF_virgo} by the red plus signs. Hereinafter this sample will be referred to as the ``Virgo slice''. We also take the volume within 3 Mpc of the centre of the cluster (defined by the position of M87 and with line-of-sight distances taken as those calculated from the procedure described in the Section \ref{sec:dist_errs}). In the literature the Virgo cluster is typically defined as a region within 5$^{\circ}$ of M87, whereas our sample extends to approximately double this radius, therefore, we will refer to sample as the ``extended Virgo cluster'' sample. This volume was chosen because over 90 per cent of the 272 ALFALFA galaxies it contains are members of the VCC, and for smaller radii the sample size decreases rapidly, while for larger radii the fraction of sources that are in the VCC plummets, likely indicating an increase in contaminants. The HIMF of this volume is also shown in Figure \ref{fig:a100_HIMF_virgo} (magenta diamonds).

The HIMFs of the two Virgo samples are compared to the rest of the $\alpha$.100a catalogue and the best fit to the entire $\alpha$.100a catalogue in Figure \ref{fig:a100_HIMF_virgo}. It can be seen by eye that all four share approximately the same low-mass slope and `knee' mass (values displayed in table \ref{tab:param_vals}). Unsurprisingly the extended Virgo cluster is much more densely packed with galaxies than the general ALFALFA sample, and the normalisation constant ($\phi_\ast$) increases from the $\alpha$.100a sample, to the Virgo slice, to the extended Virgo cluster sample. This demonstrates that, other than the normalisation, the Virgo HIMF is broadly consistent with the HIMF of the whole ALFALFA sample within 4000 \kms. It should be noted, however, that the `knee' mass is quite poorly constrained due to the low number of high HI mass objects in this small volume. Our extended Virgo sample HIMF also appears to be approximately consistent with the observations of AGES \citep[Arecibo Galaxy Environment Survey][]{Taylor+2012,Taylor+2013}, although this comparison has a number of caveats that we discuss below.

\section{Discussion}
\label{sec:discuss}

The results shown in the previous Section demonstrate three distinct trends in the HIMF of ALFALFA:
\begin{enumerate}
    \item The low-mass slope is significantly flatter in the Arecibo Fall sky direction than in the Spring sky.
    \item The `knee' mass is almost 0.2 dex lower within $\sim$60 Mpc than it is out to $\sim$200 Mpc.
    \item The shape of the HIMF in the Virgo cluster is consistent with the shape of the HIMF of ALFALFA sources at similar distances.
\end{enumerate}
The first two of these are not directly dependent on each other as the Fall sample in both the full and nearby catalogues show the flattening of the low-mass slope (table \ref{tab:param_vals}), but the Fall $\alpha$.100 sample does not show the suppressed $M_{*}$ value that is present in all the nearby samples, even though there is a low significance suggestion that it is most suppressed in the Fall $\alpha$.100a sample (however, note that there is a strong covariance between $\alpha$ and $m_{\ast}$). One natural conclusion is that these phenomena are the result of environmental dependence, as the Arecibo Spring sky faces towards the centre of the Local Supercluster, whereas the Fall sky faces a foreground void in front of PPS. It is possible that the characteristic HI mass is somewhat reduced within the Local Supercluster, and that the low-mass slope is flatter around the PPS foreground void. As the low-mass slope of the entire $\alpha$.100 sample is necessarily dominated by the galaxies detected nearby (where those on the low-mass slope are detectable), this would explain why the Spring/Fall dichotomy is present regardless of the maximum distance cut-off of a sample. Equally, as the `knee' mass of the $\alpha$.100 sample is dominated by the sources detected at $\sim$150 Mpc, where ALFALFA detects most of its $M_{*}$ galaxies, this would naturally explain why the suppressed `knee' mass is only apparent in the low redshift samples. However, before accepting this interpretation we will first demonstrate that we have ruled out a number of potential biases that could have produced similar apparent trends.

\subsection{The effect of a small volume}

An immediate criticism of these results is that the samples within 4000 \kms \ correspond to small volumes that are not cosmologically fair. This is certainly true as the present reality is that current wide-field HI surveys are only capable of measuring the low-mass slope within the Local Volume, however, there are sufficiently many low-mass objects detected in this volume to make a statistically significant detection of a discrepancy between the low-mass slopes of the HIMFs in the Spring and Fall skies. While this may or may not be representative of the behaviour of the HI population as a whole, there is presumably still a physical explanation of this discrepancy between these two starkly different environments. We will return to discuss the low-mass slope later on, and instead focus here on the shift seen in $M_{*}$ between the local and more distant volumes, and what effect the small (local) volume considered might have on the result.

The simplest bias that could cause an apparent shift in the `knee' mass would be if the volume considered were sufficiently small that too few $M_{*}$ galaxies existed in it to accurately constrain the value in that volume. To test this hypothesis we generated 3000 mock catalogues with an equivalent source density to $\alpha$.100, but with the small volume of the nearby sample. The input HIMF of the whole sample ($\alpha = -1.25$, $\log M_{*} = 9.94$) was used, and then cut at the ALFALFA 50 per cent completeness limit \citep{Haynes+2011}. The details of the method used to generate the mocks can be found in \citet{Jones+2015}. The \textit{observed} HIMF was then calculated for each mock, exactly as it would be for ALFALFA. The lowest `knee' mass found in any of the mocks was $\log M_{*} = 9.82$, and based on the distribution of all the mock values there is a vanishingly small probability of obtaining a value as extreme as $\log M_{*} = 9.76$. This demonstrates that the variance in the parameters cannot be attributed to the small number of sources in the Local Volume alone.


In the Local Volume distances are always a major concern because assuming a Hubble flow velocity field is not a viable option due to it being comparable to the magnitude of galaxy peculiar velocities, and without accurate distances, accurate masses cannot be determined. ALFALFA calculates most distances in the Local Volume based on the flow model of \citet{Masters2005}, but while this is certainly an improvement over assuming Hubble flow, it does not completely alleviate these concerns. There are however a number of observations that lend support for this method, and indicate it is not the cause of the effects we find. First of all the Virgo samples show suppressed values of $M_{*}$. The extended Virgo sample contains almost exclusively objects which have been assigned distances from the VCC (see Section \ref{sec:discuss_virgo}). These are objects that have been identified as part of the Virgo cluster and given the primary distance measured for the relevant cloud. Therefore, the fact that the values of $M_{*}$ follow the sample pattern, barring the possibility of an exceptionally different `knee' mass within Virgo, strongly suggests the distances are reliable, at least on average. Furthermore, to reconcile the difference in $M_{*}$ between the nearby and full samples would require the nearby ($v_{\mathrm{cmb}} < 4000$ \kms) distances to be systematically underestimated by about 25 per cent. The flow model ALFALFA uses gives a velocity correction that is less than 25 per cent of the CMB frame velocity in over 70 per cent of cases for the $\alpha$.100a sample, with those occurring as positive and negative corrections in approximately equal amounts. This again indicates that a sufficient systematic error could not plausibly be introduced by this method. As a final check we used two completely different methods to estimate the distances \citep[pure Hubble flow and the flow model of][]{Mould+2000} and in both cases $M_{*}$ was still suppressed relative to the full sample (see Appendix \ref{sec:alt_dist}).

\subsection{Trends on large angular scales}

\begin{figure}
\centering
\includegraphics[width=\columnwidth]{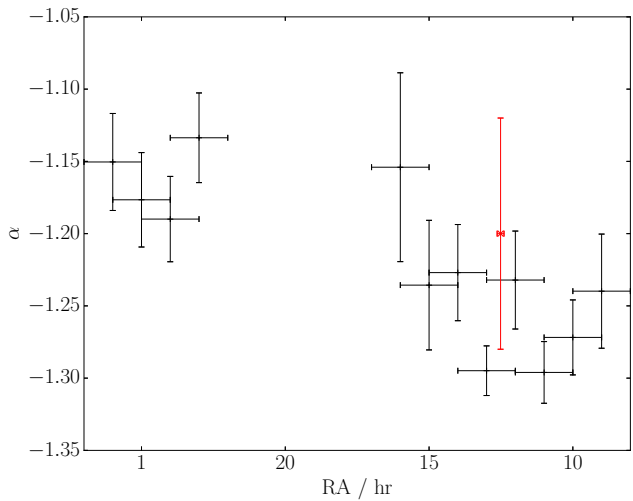}
\caption{Low-mass slope ($\alpha$) fit parameter for 2 hr wide RA bins spaced across the ALFALFA sky. The values of $\alpha$ are shown for each bin with the black error bars. The single red cross and corresponding error bars indicate the value for the extended Virgo cluster sample. There appears to be a steepening of the low-mass slope that is associated with the direction of the Virgo cluster, but extends much further in RA.}
\label{fig:alpha_RA}
\end{figure}

Figure \ref{fig:a100_HIMF_SF} shows that the low-mass slope in the Fall sky is flatter at over 3-$\sigma$ (Poisson) significance compared to the Spring sky. 
The Fall sky contains a deep void (in front of PPS), suggesting that this trend might be associated with the underdensity in the region. However, the low-mass slope of the optical luminosity function (LF) of void galaxies in the SDSS has been found to be consistent with global population \citep{Hoyle+2005,Moorman+2015}. Therefore, if the observed trend is associated with the presence of a void then the low-mass HI population would need to respond differently to underdensities than the low-mass optical population. Whether or not this is the case is unclear, but it is possible that the UV background or star formation events could preferentially ionise the HI in low-mass galaxies in voids, leading to different behaviours in the optical and HI low-mass slopes with large scale environment. To further test this hypothesis requires a population of low-mass galaxies in HI to be detected in other large scale underdensities. This may be accomplished by the next generation of HI surveys about to be carried out by Square Kilometre Array (SKA) pathfinder facilities, and will certainly be achievable for the SKA itself.

Turning our attention now to the Spring sky, the trend of the steeper low-mass slope appears to be associated with the Virgo direction, however there are not enough sources in the Virgo cluster alone to dominate the form of the HIMF over the whole Spring sky. The greater Virgo region in the ALFALFA volume is a complex knot of filaments that are in the process of falling onto the cluster, and it appears that we are looking approximately along the length of the main filament \citep{Solanes+2002,Mei+2007}. The whole structure is confined to a few hours in RA and a few tens of degrees in Dec, meaning it covers most of ALFALFA's Dec range, but a much smaller fraction of its RA range (in the Spring sky). To assess to what extent the steeper slope is associated with Virgo rather than the Spring sky as a whole, the HIMF was calculated in 2 hr wide (overlapping) bins of RA, spaced every 1 hr across the whole $\alpha$.100 sky. Figure \ref{fig:alpha_RA} shows the measured value of $\alpha$ in each RA bin. 

A sharp steepening of the low-mass slope is clearly visible centred on 12 hr (the approximate direction of the Virgo cluster), where $\alpha$ decreases around to -1.30, before rising again to nearer -1.15. In the Fall sky there is no visible trend with the same slope being consistent with all bins. Encouragingly this pattern is preserved when distances are calculated either with the \citet{Mould+2000} method or with CMB velocities, indicating that it is not an artefact of our distance estimation method.

The location of the extended Virgo cluster sample low-mass slope is also plotted in red on Figure \ref{fig:alpha_RA}. Although the error bar is large, this indicates that Virgo is not dominating the observed steep slope in this RA slice, and in fact suggests that the slope in Virgo may well be flatter than in its immediate surroundings.

The Virgo cluster itself appears not to be driving the observed steepening of the low-mass slope, yet this phenomenon is strongly associated with the RA of the cluster, which leads us to hypothesise that the filaments connected to Virgo may be driving this shift in $\alpha$, and that this could be a more general effect of a galaxy's position in LSS (specifically gas-rich filaments).

\subsection{The Virgo cluster}
\label{sec:discuss_virgo}

The Virgo cluster is known to be HI-deficient \citep[e.g.][]{Giovanelli+1985,Solanes+2001,Boselli+2006,Chung+2009,Taylor+2012} and objects falling into Virgo have been seen directly to be losing their gas \citep{Kenney+2004}, therefore, it is perhaps somewhat surprising that its HIMF appears to be generally consistent with the rest of the ALFALFA sample at similar distances (except for the normalisation). However, as is shown in Figure \ref{fig:alpha_RA} the low-mass slope of Virgo does seem to be considerably flatter than that of its immediate surroundings, even though it is consistent (at 1-$\sigma$) with the $\alpha$.100a sample as a whole. Therefore, we make the tentative suggestion that the greater Virgo region might be rich in HI, leading to a steep low-mass slope, but that the cluster instead is HI-deficient and, accordingly, has a flatter slope.

\citet{Davies+2004} studied the Virgo cluster in HI with the Jodrell Bank Lovell telescope, and concluded that there is a dearth of low-mass HI galaxies relative to the HIPASS field population \citep{Zwaan+2003}, and this result was confirmed by \citet{Gavazzi+2005} with Arecibo observation of the Virgo cluster. However, neither of these works calculated the formal HIMF because they did not weight their detections based on their completeness and sensitivity limits. More recently, the HI population detected by ALFALFA in the Virgo region was compared with AGES \citep{Taylor+2012,Taylor+2013}. The two AGES fields within Virgo covered 25 square degrees, sampling both the cluster centre and its outskirts at a sensitivity about a factor of $\sim$4 times greater than ALFALFA. They find a lower fraction of massive galaxies than ALFALFA, but point out that this is not very significant given the small area that they cover and the infrequency of such galaxies. They also find a much higher fraction of low-mass galaxies (below $\log M_{\mathrm{HI}}/\mathrm{M_{\odot}} < 7.5$) than ALFALFA. However, as no completeness limit is derived for that dataset, the formal HIMF is not calculated in those articles. This makes drawing any conclusions as to whether AGES population is or is not consistent with the low-mass slope that ALFALFA finds in Virgo problematic. 

AGES detects a considerably higher fraction of low-mass objects in Virgo than ALFALFA does, but this in itself is not surprising as it was more sensitive, and the detected fractions are in agreement in the intermediate mass range where neither sensitivity nor survey area lead to large differences \citep[see][their Figure 2]{Taylor+2013}. The lower fraction of low-mass objects detected by ALFALFA is corrected in the HIMF calculation because these object have much smaller $V_{\mathrm{eff}}$ volumes than more massive objects, but without a similar correction for the AGES Virgo sources a direct comparison is not possible. To this end we make two strong assumptions to permit such a comparison: 1) that AGES is entirely complete within the surveyed regions in the Virgo cluster (A cloud at 17 Mpc), 2) the volume surveyed is exactly the area of the rectangular fields times the assumed depth of the cluster, 2.4 Mpc \citep{Mei+2007}. With these two assumptions we can make a pseudo-HIMF from the AGES detections, which is shown in Figure \ref{fig:a100_HIMF_virgo}. This reveals that the ALFALFA and AGES datasets appear to be largely consistent within Virgo, although there is a slight suggestion that the normalisation of the AGES HIMF may be higher, which is not surprising given that the larger of their two Virgo fields targeted a particularly dense region of the cluster. It can also be seen that although AGES detects lower mass sources due to their greater sensitivity, the corrections applied to ALFALFA imply that there should be considerably more low-mass HI sources than even what AGES detected, i.e. for the lowest mass sources our first assumption appears to be invalid.

The HI-deficiency in the cluster would also be expected to suppress the `knee' mass, at least in the cluster core. Whether or not this has occurred is difficult to determine as $M_\ast$ is very poorly constrained in the extended Virgo cluster sample. However, the reader may have noted that there are no galaxies at all above an HI mass of $10^{9.8}$ \Msol \ \citep[this was also noted by][]{Kent2008}. This may seem like a significant observation, but when the $\alpha$.100a Spring HIMF is integrated starting from this mass, it shows that (based on a spherical volume of radius 3 Mpc) the extended Virgo cluster sample would only be expected to have fewer than 2 galaxies with HI masses greater than $10^{9.8}$ \Msol \ (assuming the cluster HIMF is 1.5 orders of magnitude more densely populated than the field). Thus, their absence could easily be explained by simple small number statistics. To accurately constrain the `knee' mass in the Virgo cluster requires a larger sample. Unfortunately, this is not simply a matter of sensitivity, as ALFALFA has more than sufficient sensitivity to detect any HI $M_{\ast}$ galaxies at that distance. The statistics are poor solely because a structure like the Virgo cluster does not contain many massive HI galaxies. Therefore, to confidently constrain the `knee' mass in young clusters like Virgo requires surveying multiple such clusters at a similar HI mass sensitivity to what ALFALFA achieves in Virgo. Such a survey is only feasible with the capabilities of the SKA.

\subsection{Comparison and conflicts with previous results}

\begin{figure}
    \centering
    \includegraphics[width=\columnwidth]{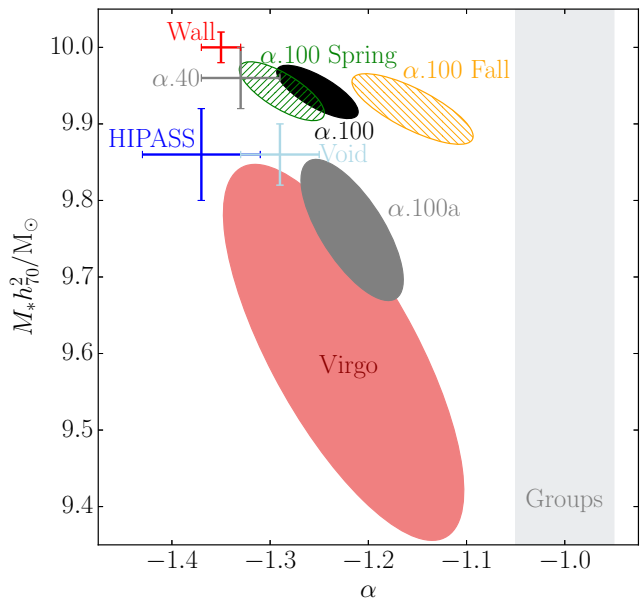}
    \caption{Comparison of various measurements of the HIMF shape parameters with 2-$\sigma$ (Poisson) error bars or ellipses. The solid black ellipse is the HIMF of the full $\alpha$.100 sample. The green (upwards hatching) ellipse is the Spring side of ALFALFA, and the orange (downward hatching) ellipse is the Fall side. Here the Virgo sample (solid light red ellipse) is the Virgo slice sample covering one hour of RA centred on the cluster out to a maximum CMB velocity of 3000 \kms. The $\alpha$.100a (solid grey) sample covers the full footprint out to 4000 \kms. The $\alpha$.40 (grey) and HIPASS (blue) points correspond to the HIMF of the 40 per cent ALFALFA catalogue \citep{Martin+2010} and the HIPASS catalogue \citep{Zwaan+2005}. The Void (light blue) and Wall (red) samples are the results of \citet{Moorman+2014}, based on ALFALFA 40 per cent. The grey band on the right side of the plot shows approximately the typical range of slopes found for individual groups of galaxies. For more details on the $\alpha$.100 samples refer to Table \ref{tab:param_vals}.}
    \label{fig:all_HIMF_comp}
\end{figure}

Figure \ref{fig:all_HIMF_comp} shows Schechter function shape parameters (the low-mass slope and the `knee' mass) for the sub-regions discussed in this paper, along with several previous measurements of the HIMF and its variation with environment. In this Section we discuss each of these comparisons and their possible interpretations in turn.

\subsubsection{Environmental dependence}

The fact that our results suggest a large scale environmental dependence of the low-mass slope seems to be in conflict with \citet{Moorman+2014} and \citet{Jones+2016b}, both of which studied the environmental dependence of the HIMF in ALFALFA (on large and small scales) and concluded that the `knee' mass does vary on the order of 0.1 dex with environment, but neither found convincing evidence for a change in the low-mass slope. However, there are a number of plausible explanations for why this trend has not been seen before. 

\citet{Moorman+2014} defined their void and wall environments based on SDSS spectroscopy. These two measurements are shown by the light blue and red error bars in Figure \ref{fig:all_HIMF_comp}, and labelled as ``Void" and ``Wall". That work concluded that the `knee' mass was lower within voids than in walls, and found a tentative flattening of the low mass slope at low significance. As there is only substantial overlap between ALFALFA and SDSS spectroscopy in the Arecibo Spring sky this would immediately prevent that study from finding the large dichotomy between the Spring and Fall skies that we observe.

\citet{Jones+2016b} focused on a smaller scale definition of environment based on density of neighbours in SDSS DR8 \citep{Aihara+2011} or 2MRS \citep[2MASS Redshift Survey,][]{Huchra+2012} around ALFALFA sources. That work did incorporate the Fall portion of $\alpha$.70 in some of their analysis (when using neighbour densities from 2MRS, which is all sky), but the definition of environment was too small scale to identify the dichotomy found here, and while it may have been large enough angular scale to find the trend associated with the Virgo direction, it was not focused on just the local structures. Therefore, its null result is unsurprising. 
\citet{Jones+2016b} also concluded that the shift in the value of $M_{*}$ was associated with local environment ($\sim$2 Mpc scales), with it disappearing when environment was defined on a larger scale ($\sim$10 Mpc). Again, this is in apparent conflict with our present results (lower $M_{*}$ in the Local Supercluster).

Knowing the results of this work we returned to the environment (neighbour density) quartiles calculated for $\alpha$.70 in \citet{Jones+2016b} and determined whether $\alpha$.100 sources fell within voids or walls based on the \citet{Moorman+2014} analysis. When split into RA bins across the Spring region of ALFALFA the mean environment quartile shows a similar pattern (although inverted) to that of Figure \ref{fig:alpha_RA}, with the majority of the sources in the immediately vicinity of Virgo being classified as in high density environments, but the average environment quartile decreases towards the edges of the Spring sky. The fraction of sources classified as in voids also climbs either side of Virgo, with the majority of ALFALFA's low-mass sources to the west of Virgo lying in voids and almost half at the eastern edge of the Spring sky. This pattern suggests that the variation in the low-mass slope is due to environment, with the steepest slopes being in high density environments around Virgo (but not in the cluster itself), and the flatter slopes in lower density environments. This also raises the possibility that rather than the gas-rich regions connected to the Virgo cluster driving the slope to be steeper in that region, it may be the low density environments elsewhere that are driving it to be flatter, or perhaps both. A deeper survey than ALFALFA is needed to further explore this dependence in the low-mass slope over a larger volume.

We also found that within the nearby volume the majority of the $M_{\ast}$ galaxies were classified as being in the lower half of neighbour densities (compared to the full $\alpha$.70 sample), and that more than twice as many were classified as lying in voids than in walls (compared to a ratio of 2:1 in the opposite sense for the full sample). This would then imply agreement with the findings of \citet{Moorman+2014} and \citet{Jones+2016b} that the `knee' mass is decreased in low density environments, and that that is causing the observed suppression of the local value. 

While all these findings appear to form a coherent picture, caution is advised because this agreement could still be circumstantial and not indicative of the root cause of these variations.

\subsubsection{HIPASS}

The lower `knee' mass that we observe in the local Universe conveniently explains the apparent discrepancy in `knee' masses measured by HIPASS and ALFALFA \citep{Zwaan+2005,Martin+2010}. The `knee' mass that we calculate is 0.08 dex higher than the HIPASS value \citep[0.1 dex compared to][]{Martin+2010}. Although HIPASS's nominal volume is larger than ALFALFA's, due to its superior sensitivity the volume in which ALFALFA can detect a galaxy of $\log M_{\mathrm{HI}} = 10$ is over 6 times greater than that available to HIPASS \citep[estimated based on 50 per cent completeness limits,][]{Zwaan+2004,Haynes+2011}. This, combined with our finding of a lower local value of $M_{*}$, suggests that HIPASS's lower value of $M_{*}$ (see Figure \ref{fig:all_HIMF_comp}), was caused by their `knee' mass determination being dominated by the local region where its value appears to be low.

HIPASS also found a steeper low-mass slope than any of the ALFALFA samples. We have seen that the value of $\alpha$ can change on large angular scales, so there remains the definite possibility that the low-mass slope is intrinsically steeper in the Southern sky. However, as HIPASS used pure Hubble flow distances (rather than using a flow model) and implemented their completeness limit slightly differently to ALFALFA, caution is advised when interpreting this difference.

\subsubsection{Groups}

Studies of the HIMF in individual groups \citep[e.g.][]{Verheijen+2001,Kovac+2005,Freeland+2009,Kilborn+2009,Pisano+2011,Westmeier+2017} have generally found that they have approximately flat ($\alpha = -1$) low-mass slopes, with the notable exceptions of \citet{Stierwalt+2009} and \citet{Davies+2011}, both of which found very steep ($\alpha < -1.4$) low-mass slopes in the Leo region and across various groups within the AGES footprint, respectively. These results are apparently in tension with those of ALFALFA and HIPASS, which measured a steeper slope ($\alpha \sim -1.3$) and do not find a flattening of the slope in higher density regions \citep{Zwaan+2005,Martin+2010,Moorman+2014,Jones+2016b}. 

The large scale shifts in the value of $\alpha$ that we have found highlight that it is important for HI studies of individual groups to consider the larger scale environment of groups when comparing their HIMFs to those of wide-field surveys. Although ALFALFA does not cover the sky areas where most of these group studies have been carried out, it is clear from our results that large scale environmental changes can cause substantial shifts in $\alpha$, meaning that the position of a group within LSS may be comparably important as the difference in environment between the field and groups. Having said this, none of the sub-regions or previous results plotted in Figure \ref{fig:all_HIMF_comp} enter the grey band that approximately represents the findings in galaxy groups, so this does not appear to be a complete explanation of the apparent tension. 

There have been several recent indications that group galaxies are ``pre-processed" \citep[e.g.][]{Hess+2013,Odekon+2016,Brown+2017}, losing a significant amount of their HI upon joining groups. While this pre-processing could be an explanation for the flatter low-mass slope found in galaxy groups, without reference to the larger scale environment, it does not explain why a transitional HIMF has not been identified with ALFALFA or HIPASS. In addition, low stellar mass group centrals may actually have more HI than their counterparts in the field \citep{Janowiecki+2017}, which suggests that the apparent amount of processing is probably also a function of group size and potentially position relative to the cosmic web.

An additional complicating factor is that most groups have been studied using interferometric data
(again \cite{Stierwalt+2009,Davies+2011} are exceptions, but so are \cite{Pisano+2011,Westmeier+2017}), whereas HIPASS and ALFALFA are both single dish surveys. The completeness limit is extremely important when calculating the HIMF and interferometers suffer from the additional complication that there is a surface brightness limit as well as an integrated flux limit. This is irrelevant in almost all cases for single dish telescopes as they seldom resolve any but the most nearby sources. If not adequately accounted for this effect could lead to an under counting of low surface brightness sources and a bias in the low mass slope. Furthermore, the completeness limit of HI surveys also depends on the source velocity width, because a spectrum can be smoothed to enhance signal-to-noise. This means that some sources of a given HI integrated flux may be detected, but other sources with the same flux may not be detectable because their velocity width is too wide and their flux is spread across too many channels. In Appendix \ref{sec:wid_corr} we argue that in the case of groups a correction for this effect should always be applied, however, it typically is not, and this can lead to an underestimation of the slope. 

\citet{Minchin2017} argued that by combining flat Schechter function sub-HIMFs (HIMFs of sub-populations or groups) of different `knee' masses the global HIMF shape could be recovered, which would resolve the conflict as groups could have flat slopes provided there was the appropriate distribution of `knee' masses in the sub-HIMFs across all environments. While this scenario is difficult to disprove, the vast majority of ALFALFA sources are not members of groups, but are centrals in the field \citep{Guo+2017}, thus, this approach would amount to summing together a different (flat) sub-HIMF for each galaxy, which prompts the question: what do we mean by a MF and can an individual galaxy have an associated MF? This, however, raises an important point, which is that the ALFALFA population principally probes the HI content of centrals, whereas studies of individual groups probe the HI content of the satellites (as each group can only have one central). Therefore, as they detect distinct populations it is perhaps unsurprising that the two types of HI datasets have a disconnect in the observed HIMF slopes and that a transition has not been identified in ALFALFA.

\subsection{Prospects for future HI surveys}

Over the next 5 to 10 years there will be a slew of new blind HI surveys that will be performed by SKA-pathfinder facilities \citep[e.g.][]{Duffy+2012c,Giovanelli+2016}. The direct successors to HIPASS and ALFALFA will be WALLABY (Widefield ASKAP L-band Legacy All-sky Blind surveY) in the Southern hemisphere and WNSHS (Westerbork Northern Sky HI Survey) in the Northern hemisphere. Together these HI surveys will cover the entire sky and detect an order of magnitude more galaxies than ALFALFA. Another class of surveys will also be carried out that perform deep integrations of limited fields. These surveys will come in two varieties: 1) mapping of small fields such as the Westerbork Medium Deep survey and the DINGO survey \citep[Deep Investigation of Neutral Gas Origins,][]{Meyer2009}, which will detect a comparable number of sources to ALFALFA, but out to $z \sim 0.25$, and 2) extremely deep single pointing surveys that aim to detect HI galaxies out to a redshift of order unity, such as LADUMA \citep[Looking At the Distant Universe with MeerKAT,][]{Holwerda+2012} and the ongoing CHILES \citep[COSMOS HI Large Extragalactic Survey,][]{Fernandez+2013}.

The results of this paper suggest that these surveys will contribute to the study of the HIMF in two key ways. First, the low-mass slope will be measurable in the field outside the Local Volume for the first time. This will give the first cosmologically fair account of how numerous low-mass, gas-rich galaxies really are at $z = 0$. And second, the `knee' mass will become measurable beyond $z \approx 0$. Due to the volume of ALFALFA these next generation surveys are unlikely to find a significantly different `knee' mass, however, they will have the depth and source counts to begin to place constraints on its evolution as a function of $z$. 

While the redshift range covered by SKA-precursor surveys will far exceed what is possible with single dish telescopes \citep[due to the increase in source confusion with redshift,][]{Jones+2015,Elson+2016}, interferometric surveys suffer a degradation in their nominal sensitivity for nearby objects which can be resolved over many synthesised beams (suppressing the signal-to-noise like the square root of the number of beams). This is seldom a problem for single dish telescopes as even the largest dishes have beams that a few arcmin across at 21 cm wavelengths. Furthermore, if outfitted with multi-beam receivers capable of forming tens of beams on the sky, the survey mapping speed of the largest existing single dish telescopes could effectively match that of the SKA-pathfinder interferometers (at equivalent nominal sensitivity), making them ideal instruments to detect the very lowest HI mass, and most diffuse, galaxies in the nearby Universe and address the question of what is the threshold mass to form a galaxy.

\section{Conclusions}
\label{sec:conclude}

We have measured the HIMF of the full ALFALFA survey, confirming it is well fit by a Schechter function with parameters $\alpha = -1.25 \pm 0.02 \pm 0.03$, $m_{*} = 9.94 \pm 0.01 \pm 0.04$, and $\phi_{*} = (4.5 \pm 0.2 \pm 0.4) \times 10^{-3} \; \mathrm{Mpc^{-3} \, dex^{-1}}$, where the first error quoted is the estimated random uncertainty due to Poisson counting errors, flux measurement errors, and peculiar velocities, the latter is the estimate of the systematic uncertainty due to boundary effects, sample variance, choice of flow model, the absolute flux scale, and bias inherent to the $V_{\mathrm{eff}}$ (or 2DSWML) method. This is the most precise measurement of the HIMF, and the most complete accounting of its uncertainties, to date.

If estimates of the impact of cosmic variance are also included then there is little impact on the systematic uncertainties in the `knee mass', which rise to 0.05 dex, but the low-mass slope becomes highly uncertain, with an estimated systematic error of $\sim$0.1. This indicates that ALFALFA has made a cosmologically fair measurement of the HIMF `knee' mass, but although it has a very precise measurement of the low-mass slope, its value only reflects that of the local Universe. Due to this large systematic uncertainty in $\alpha$ the measurement of the global HI content of the $z=0$ Universe (corrected for self-absorption) also has a large systematic uncertainty, $\Omega_{\mathrm{HI}} = (3.9 \pm 0.1 \pm 0.6) \times 10^{-4}$.

We also investigated differences in the ALFALFA HIMF between the Arecibo Spring and Fall skies, and the local ($v < 4000$ \kms) and full volumes ($v < 15000$ \kms). We find a clear dichotomy in the low-mass slope of the Spring and Fall skies that was missed by previous studies of the environmental dependence of the ALFALFA population because they focused on the Spring sky only or smaller scale definitions of environment. 

The Spring sky is dominated by the Virgo cluster, a very overdense region, whereas, at a similar distance, the Fall sky contains a deep void. This coincident shift in environment and low-mass slope strongly suggests the two are connected. Furthermore, the steepening of the slope appears to be associated with the direction of the Virgo cluster, but the cluster itself does not appear to be the dominant component driving a steeper slope. We therefore hypothesise that the steeper slope may be associated with the gas-rich filaments that are connected to Virgo and feeding the growth of the cluster, or that the flattening of the slope away from the cluster may be due to the impact of low density environments. To test these hypotheses will require much deeper surveys than ALFALFA that would be capable of detecting equivalent phenomena, should they exist, in and around other clusters and voids. 

The existence of shifts in $\alpha$ on large angular scales also indicate that caution should be exercised when comparing the results of studies of individual galaxy groups to wide-field surveys, as the HI-content of their constituent galaxies may be impacted by the group's presence within a larger structure as well as membership of the group itself. We have also discussed several methodological considerations that are relevant to such a comparison, and note that studies of groups detect almost exclusively satellites, while wide-area blind HI surveys consist almost entirely of field centrals.

The other shape parameter of the HIMF, `knee' mass, appears relatively independent of direction on the sky, indicating that the ALFALFA volume is deep enough to give a cosmologically fair representation of this parameter at $z=0$. It does however, appear to have a somewhat lower value in the local Universe (compared to the whole ALFALFA sample), which conveniently explains the lower value found by HIPASS \citep{Zwaan+2005}, as HIPASS was a shallower survey than ALFALFA. 

Upcoming surveys with SKA-pathfinders will be capable of addressing these unresolved issues and will create a much tighter constraint on the HI population as a function of environment, and eventually, redshift. As simulations continue to include more and finer detail gas physics, these surveys will be vital to pointing those efforts in the correct direction. However, the largest existing single dish radio-telescopes may turn out to be the best suited instruments to addressing questions concerning the lowest mass and surface density galaxies in the Universe.

\section*{Acknowledgements}

The authors acknowledge the work of the entire ALFALFA collaboration in observing, flagging, and extracting the catalogue of galaxies that this paper makes use of. The ALFALFA team at Cornell is supported by NSF grants AST-0607007, AST-1107390 and AST-1714828, and by grants from the Brinson Foundation. MGJ also acknowledges support from the grant AYA2015-65973-C3-1-R (MINECO/FEDER, UE). We also acknowledge helpful comments from A. Robotham concerning the satellite and central mass functions. We thank the referee for their thorough reading of the manuscript.

This work made use of SDSS data products. Funding for the Sloan Digital Sky Survey IV has been provided by the Alfred P. Sloan Foundation, the U.S. Department of Energy Office of Science, and the Participating Institutions. SDSS acknowledges support and resources from the Center for High-Performance Computing at the University of Utah. The SDSS web site is www.sdss.org.

\bibliography{refs}

\appendix

\section{ALFALFA and HIPASS volume calculations}
\label{app:vol}

The calculations in the introduction showing that the accessible volume for ALFALFA is larger for both the low-mass slope and the `knee' mass are somewhat counter intuitive because the total volume of HIPASS is larger than ALFALFA, and HIPASS continues to detect sources almost to the edge of its bandwidth. When comparing to \citep{Zwaan+2005} (their Figure 2) it can be seen that $M_{*}$ galaxies are detected beyond 75 Mpc, the maximum detection distance calculated for galaxies of $\log M_{\mathrm{HI}}/\mathrm{M_{\odot}} = 10$ and velocity width of 300 \kms \ using equation 4 of \citet{Zwaan+2004}. However, these detections generally have narrower velocity widths than is typical for such galaxies, which increases their peak flux, making them detectable out to greater distances. The same would be true for ALFALFA, which would cause its volume to grow at a faster rate than HIPASS's (as volume grows like $r^{3}$ and the detection distance is greater for ALFALFA), but the outer edge of ALFALFA is determined by a band of heavy RFI, which means the volume cannot grow much greater. This creates an odd situation where which survey has a greater volume depends on the mass and the velocity width of the galaxies in question. For galaxies of $\log M_{\mathrm{HI}}/\mathrm{M_{\odot}} = 10$ with velocity widths greater than $\sim$100 \kms \ ALFALFA has the larger volume, but for narrower sources HIPASS has the larger volume. As most $M_{*}$ galaxies have broader velocity widths than this we conclude that ALFALFA has the larger volume for probing the `knee' mass.

Essentially the same point can also be made in a simpler way, without explicit reference to the survey detection limits, but in doing so hides the details of what is happening. The median distance to a HIPASS $M_{*}$ galaxy is approximately 50 Mpc, whereas for ALFALFA it is approximately 150 Mpc. Therefore, if the survey areas were the same, the volume accessible in ALFALFA would be 27 times larger for the typical $M_{*}$ galaxy. As the survey area of HIPASS is less than 27 times that of ALFALFA, the accessible volume for $M_{*}$ galaxies must be larger for ALFALFA.

For the low-mass slope the situation is much simpler because these sources in both HIPASS and ALFALFA are detected well away from the outer edge of the bandwidth and the band of RFI that truncates ALFALFA. Thus, lowering the velocity width used to calculate the maximum detection distance just causes the accessible ALFALFA volume to grow faster than the accessible HIPASS volume, as would be naively expected.

\section{Alternative distance estimates}
\label{sec:alt_dist}

\begin{figure}
\centering
\includegraphics[width=\columnwidth]{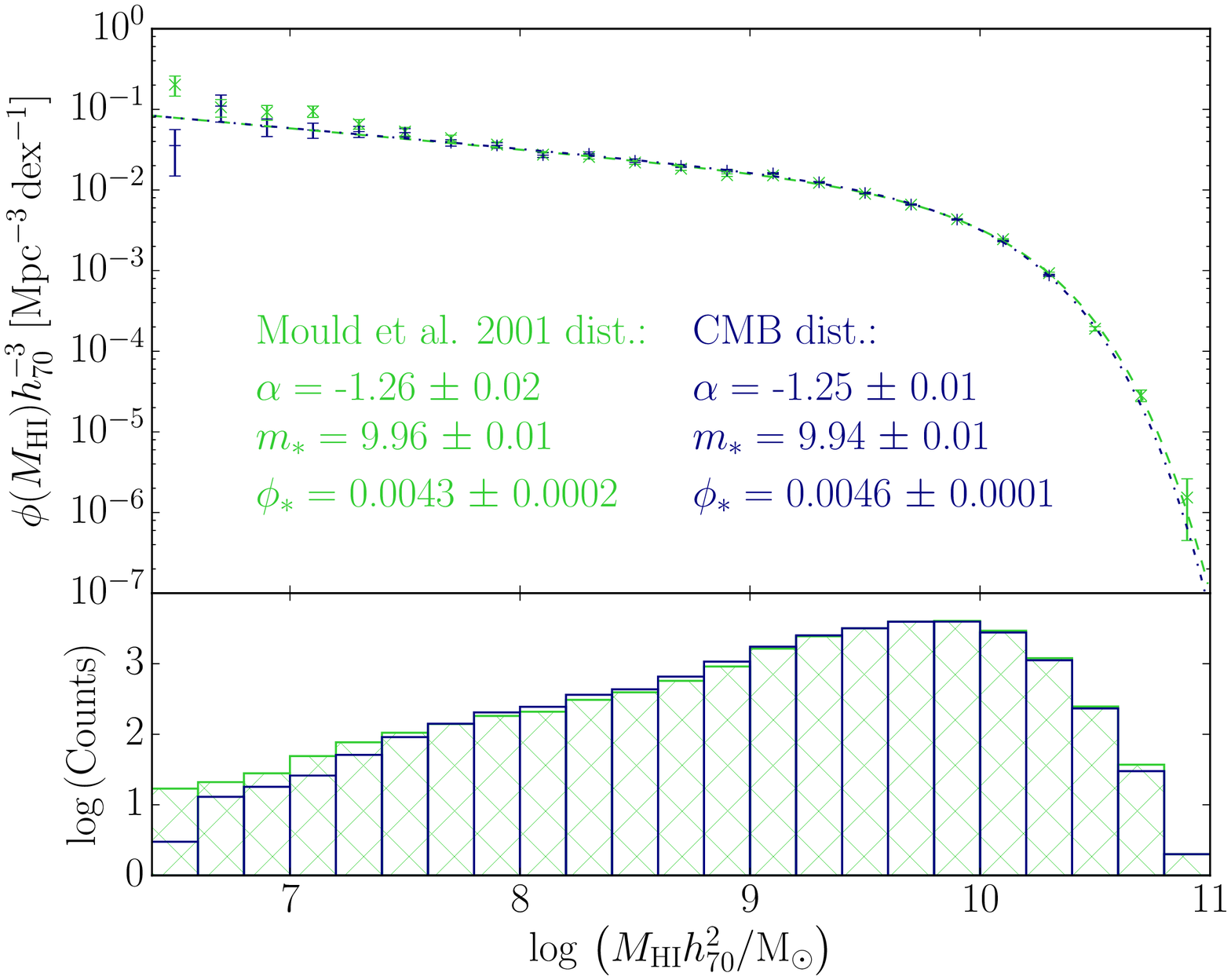}
\caption{The HIMFs of the $\alpha$.100 sample for \citet{Mould+2000} method distances (light green crosses and cross hatched bars) and distances calculated directly from the CMB velocities (dark blue pluses and unfilled bars).}
\label{fig:HIMF_alt_dists}
\end{figure}

\begin{table*}
\centering
\caption{As in Table \ref{tab:param_vals}, but for samples using the alternative distance estimates.}
\label{tab:alt_dist_params}
\begin{tabular}{| l | c | c | c | c | c |}
\hline \hline
Sample              & $v_{\mathrm{max}} / \mathrm{km\,s^{-1}}$ & $N_{\mathrm{gal}}$ & $\alpha$ & $m_{*}$ & $\phi_{*} / \mathrm{Mpc^{-3} \, dex^{-1}}$          \\  
\hline
$\alpha.100$ (Hubble)         & 15000                                            & 22815                      & $-1.25 \pm 0.01$  & $9.94 \pm 0.01$  & $4.6 \pm 0.1 \times 10^{-3}$ \\
$\alpha.100$ Spring (Hubble)  & 15000                                            & 14392                      & $-1.27 \pm 0.01$  & $9.94 \pm 0.01$  & $5.0 \pm 0.1 \times 10^{-3}$ \\
$\alpha.100$ Fall (Hubble)    & 15000                                            & 8423                       & $-1.23 \pm 0.01$  & $9.94 \pm 0.01$  & $3.9 \pm 0.1 \times 10^{-3}$ \\
$\alpha.100a$ (Hubble)        & 4000                                             & 3804                       & $-1.22 \pm 0.02$  & $9.74 \pm 0.04$  & $6.4 \pm 0.6 \times 10^{-3}$ \\
$\alpha.100$ (Mould et al.)   & 15000                                            & 22693                      & $-1.26 \pm 0.02$  & $9.96 \pm 0.01$  & $4.3 \pm 0.2\times 10^{-3}$ \\
$\alpha.100$ Spring (Mould et al.)   & 15000                                     & 14256                      & $-1.30 \pm 0.02$  & $9.97 \pm 0.01$  & $4.5 \pm 0.2\times 10^{-3}$ \\
$\alpha.100$ Fall (Mould et al.)     & 15000                                     & 8437                       & $-1.14 \pm 0.02$  & $9.92 \pm 0.02$  & $4.4 \pm 0.2\times 10^{-3}$ \\
$\alpha.100a$ (Mould et al.)  & 4000                                             & 3472                       & $-1.29 \pm 0.02$  & $9.85 \pm 0.05$  & $4.3 \pm 0.5\times 10^{-3}$ \\
\hline
\end{tabular}
\end{table*}

One critisim of this work is that all masses are reliant on distances, which are not measured directly by ALFALFA. Our flow model \citep{Masters2005} makes corrections to the velocities of nearby objects and then uses Hubble's law to calculate distance estimates. There are also additional corrections for galaxies assigned to groups or the Virgo cluster (described in Section \ref{sec:dist_errs}), or those which have primary or secondary distance measurements in the literature. If the distances that this model produced are systematically bias then some of our finding could be due, in part, to this bias.

To address this issue we have recalculated the distances to all ALFALFA sources using two alternative methods. The first is our own implementation of the \citet{Mould+2000} flow model which included simple spherical infall around Virgo, Shapley and the Great Attractor. The second is to simply use the CMB velocity of the sources with no corrections at all. 
The resulting HIMFs for these two distance estimation schemes are shown in Figure \ref{fig:HIMF_alt_dists}. The HIMFs are almost indistinguishable from each other and the fit parameters are consistent within 1-$\sigma$ errors of those of the HIMF based on ALFALFA's flow model.

It is also noticeable that the CMB distances lead to many more nearby sources being excluded (bottom panel of Figure \ref{fig:HIMF_alt_dists}). Part of the reason for this is because it is not uncommon for nearby sources with large peculiar velocities (often in the direction of Virgo) to have negative CMB velocities. This means they are immediately excluded. The \citet{Mould+2000} has the opposite effect and overestimates the number of very low mass sources (relative to ALFALFA's flow model), which indicates it is calculating smaller distances for some of the nearest sources. Despite this difference, the sample with CMB distances retains more objects overall. This is because at large distances the \citet{Mould+2000} model slightly over-estimates the velocities of sources and therefore places them at larger distances than the pure Hubble flow model, which leads to more being cut by the maximum distance limit that we set ($v_{\mathrm{max}}/H_{0}$). This difference is not immediately apparent from Figure \ref{fig:HIMF_alt_dists} due to the logarithmic scale, but the counts in the intermediate mass bins are systematically higher for the Hubble flow sample than the \citet{Mould+2000} flow model sample.

The fit parameters for the full list of sub-samples is shown in Table \ref{tab:alt_dist_params}.

\section{Width correction factor}
\label{sec:wid_corr}

In the calculation of the $\alpha$.40 HIMF \citep{Martin+2010} a correction for sources that are undetected due to their velocity widths being too broad was applied, however, we make no correction here because we do not set a minimum distance cut on the sample. The reason why the minimum distance limit is the determining factor is explained below, by means of an example.

Imagine that at a given distance $D_{1}$ a galaxy of HI mass $M_{\mathrm{1}}$ is above the completeness limit only if it has a narrow velocity width, but is undetectable if it has a broad velocity width. In the case of a large area survey there will always be (except for the most extremely nearby sources) nearer sources of the same mass, that, because they are nearer, will be detectable even if they have very broad velocity widths. This means that the 2DSWML algorithm can correctly approximate the form of the mass-width function (the 2D distribution of intrinsic HI masses and velocity widths). Therefore, those sources of mass $M_{\mathrm{1}}$ which were undetected at $D_{1}$ due to their large velocity widths are automatically accounted for by the estimates of $V_{\mathrm{eff}}$ for those sources of mass $M_{\mathrm{1}}$ that were detected.

However, in the case where a minimum distance limit is set this line of reasoning no longer applies because there are no galaxies in the foreground to act to correct the $V_{\mathrm{eff}}$ estimates. Note that this also applies for the $V_{\mathrm{max}}$ method, with the additional point that this method doesn't account for the problem in the first place. As a group is, by definition, confined to a small region there is always a minimum distance limit applied and thus a width correction factor is necessary. This correction should take the form of an up-weighting of each source based on the fraction of sources of that mass that would be undetectable at each distance. This requires the mass-conditional width function to be measured or assumed for all masses.

When a lower distance limit is set and a correction is not applied we have invariably found that it acts to flatten the low mass slope by suppressing the lowest mass bins. This effect is just visible in the extended Virgo cluster HIMF plotted in Figure \ref{fig:a100_HIMF_virgo}. This was not corrected for as it only appears to strongly effect the first bin (probably due to the proximity of Virgo) and its value is highly uncertain anyway.

\section{Boundary vertices}
\label{sec:boundary}

The Tables \ref{tab:S_bound}, \ref{tab:F_bound}, \ref{tab:S_bound_st}, and \ref{tab:F_bound_st} show the coordinates of the boundary vertices of the Spring and Fall samples, and the strict Spring and Fall samples respectively.

\begin{table}
\centering
\caption{Vertices of the boundary of the Spring sample.}
\label{tab:S_bound}
\begin{tabular}{ccc}
\hline \hline
Dec range {[}deg{]} & RA min {[}hr{]} & RA max {[}hr{]} \\
\hline
0-16      & 7.7    & 16.5   \\
16-18     & 7.7    & 16.0   \\
18-20     & 8.7    & 15.4   \\
20-24     & 9.4    & 15.4   \\
24-30     & 7.6    & 16.5   \\
30-32     & 8.5    & 16.0   \\
32-36     & 9.5    & 15.5   \\
\hline
\end{tabular}
\end{table}

\begin{table}
\centering
\caption{Vertices of the boundary of the Fall sample.}
\label{tab:F_bound}
\begin{tabular}{ccc}
\hline \hline
Dec range {[}deg{]} & RA min {[}hr{]} & RA max {[}hr{]} \\
\hline
0-2                 & 22.0            & 3.0             \\
2-6                 & 22.5            & 3.0             \\
6-10                & 22.0            & 3.0             \\
10-14               & 22.0            & 2.5             \\
14-36               & 22.0            & 3.0             \\
\hline
\end{tabular}
\end{table}

\begin{table}
\centering
\caption{Vertices of the boundary of the Spring strict sample.}
\label{tab:S_bound_st}
\begin{tabular}{ccc}
\hline \hline
Dec range {[}deg{]} & RA min {[}hr{]} & RA max {[}hr{]} \\
\hline
0.5-15.5  & 8.0    & 16.0   \\
15.5-24.5 & 10.0   & 15.0   \\
24.5-29.5 & 8.0    & 16.0   \\
29.5-35.5 & 10.0   & 15.0   \\
\hline
\end{tabular}
\end{table}

\begin{table}
\centering
\caption{Vertices of the boundary of the Fall strict sample.}
\label{tab:F_bound_st}
\begin{tabular}{ccc}
\hline \hline
Dec range {[}deg{]} & RA min {[}hr{]} & RA max {[}hr{]} \\
\hline
0.5-35.5  & 22.5   & 2.5   \\
\hline
\end{tabular}
\end{table}

\end{document}